\documentclass[12pt]{article}
 \usepackage{ifpdf}
 \usepackage{epsfig}
  \usepackage[latin1]{inputenc}
 \usepackage{amsfonts}
 \usepackage{amsthm}
 \usepackage{amssymb, latexsym, amsmath}
 \usepackage{amsbsy}
\usepackage{amstext}
\usepackage{amsopn}
\usepackage{amsgen}
\usepackage[dvips]{color}
 \usepackage{graphicx}
  \usepackage{color}
   \usepackage{wasysym}
   \usepackage{deluxetable}
  \usepackage{verbatim}

 \newcommand{\inc}{{\it i}}

 \newcommand{\be}{\begin{equation}}
 \newcommand{\ee}{\end{equation}}
 \newcommand{\ba}{\begin{eqnarray}}
 \newcommand{\ea}{\end{eqnarray}}
 \newcommand{\bs}{\begin{subequations}}
 \newcommand{\es}{\end{subequations}}

 \newcommand{\erbold}{\mbox{{\boldmath $\vec r$}}}
 \newcommand{\rbold}{\mbox{{\boldmath $\vec r$}}}

  \newcommand{\ddoterbold}{\ddot{\vec{\boldmath\mbox{{${r}$}}}}}

  \newcommand{\eRbold}{\mbox{{\boldmath $\vec{R}$}}}
  \newcommand{\Rbold}{\mbox{{\boldmath $\vec{R}$}}}

\begin{document}
  \title{
     ${{~}^{^{^{
          Published~in:~
     ~Celestial~Mechanics~and~Dynamical~Astronomy
        \,,~Vol.~129\,,~pp.~177\,-\,214\,~(2017)
                  }}}}$\\
 {\Large{\textbf{{{
 Tides in a body librating about a spin-orbit resonance.\\
 Generalisation of the Darwin-Kaula theory}
 ~\\
 ~\\}
            }}}}
 \author{
                                     {\Large{Julien Frouard}}\\
                                     {\small{US Naval Observatory, Washington DC 20392}}\\
                                     {\small{e-mail: ~frouardjulien$\,$@$\,$gmail.com~$\,$
                                     }}\\
                                     \vspace{1mm}
                                      ~\\
                                      {\Large{and}}\\
                                      ~\\
                                     {\Large{Michael Efroimsky}}\\
                                     {\small{US Naval Observatory, Washington DC 20392}}\\
                                     {\small{e-mail: ~michael.efroimsky$\,$@$\,$navy.mil~$\,$}}\\
 }
     \date{}

 \maketitle

 \begin{abstract}
 The Darwin-Kaula theory of bodily tides is intended for celestial bodies rotating without libration. We demonstrate that this theory, in its customary form, is inapplicable to a librating body. Specifically, in the presence of libration in longitude, the actual spectrum of Fourier tidal modes differs from the conventional spectrum rendered by the Darwin-Kaula theory for a non-librating celestial object. This necessitates derivation of formulae for the tidal torque and the tidal heating rate, that are applicable under libration.

 We derive the tidal spectrum for longitudinal forced libration with one and two main frequencies, generalisation to more main frequencies being straightforward. (By main frequencies we understand those emerging due to the triaxiality of the librating body.) Separately, we consider a case of free libration at one frequency (once again, generalisation to more frequencies being straightforward).

 We also calculate the tidal torque. This torque provides correction to the triaxiality-caused physical libration. Our theory is not self-consistent: we assume that the tidal torque is much smaller than the permanent-triaxiality-caused torque; so the additional libration due to tides is much weaker than the main libration due to the permanent triaxiality.

 Finally, we calculate the tidal dissipation rate in a body experiencing forced libration at the main mode, or free libration at one frequency, or superimposed forced and free librations.
 \end{abstract}

 \section{Motivation and plan}

 A comprehensive study of bodily tides in the vicinity of spin-orbit resonances was provided in Efroimsky (2012$\,$a,$\,$b). Those works concentrated on situations where physical libration was negligible. However, in many actual situations of planetary and satellite dynamics, libration plays a key role. Sometimes it can provide a considerable (even leading) input into the tidal heating. In such situations, physical libration may add to the temperature of the librating object.

 In this article, we demonstrate that the spectrum of the Fourier modes of longitudinal-libration-caused tides differs from the conventional spectrum of the tidal modes rendered by the Darwin-Kaula theory. This will explain why the conventional formulae from the Darwin-Kaula theory can be used only for uniform rotation. When a body is librating, the Darwin-Kaula theory in its customary form does not take care of an additional component of the tides, the one caused by physical libration in longitude.

 We derive the spectrum of the tidal Fourier modes in a body librating in longitude, with a low obliquity; and we present an expression for the polar tidal torque acting on such a body. This torque damps free librations, and generates a small correction to forced physical libration.

 Finally, we calculate the tidal dissipation rate due to libration. It turns out that for a sufficiently large dynamical triaxiality of the librating rotator the libration-caused additional tidal heating is considerable and can even exceed the regular (unrelated to libration) tidal dissipation rate.

 \section{Forced librations in longitude\label{section2}}

 We consider an extended body of a mean radius $\,R\,$, mass $\,M\,$, and the principal moments of inertia $\,A<B<C\,$. The body is spinning about its major-inertia axis (the one related to the maximal moment of inertia $\,C\,$), and is captured into one of the spin-orbit resonances with a perturber of mass $\,M^*\,$. The body is trying to align its long (minimal-inertia) axis with the instantaneous direction to the perturber. Therefore, on eccentric orbits, libration is imposed upon the regular spin.$\,$\footnote{~Libration
  can, in principle, be developed also by a body with no triaxiality, provided the tides are intensive. In this case, libration is caused by the alternating components of the tidal torque (Ferraz-Mello 2015, Fig 2). In our treatment, however, we assume that the constant triaxiality is much larger than that caused by the tides (and, accordingly, that the torque generated by the permanent triaxiality is much stronger than the tidal torque).}
 Below we concentrate on libration in longitude. Libration in latitude is neglected in our developments, which is permissible for a small obliquity.

 \subsection{The setting\label{section2.1}}

\begin{figure}[h]
 \begin{minipage}{135mm}
 \includegraphics[width=135mm]{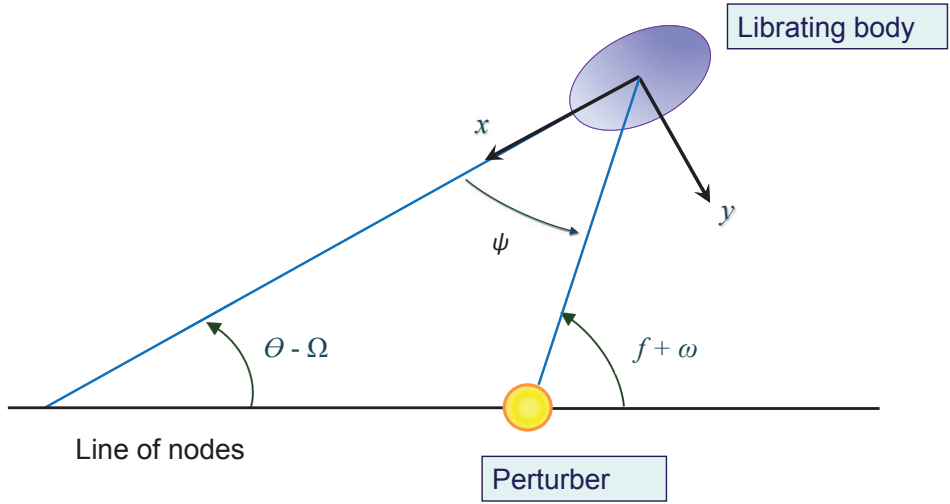}
  \end{minipage}
 \caption{\small{The principal axes $\,x\,$ and $\,y\,$ of the extended body relate to the minimal and middle moments of inertia, correspondingly. The horizontal line is that of nodes. As ever, $\,f\,$ is the true anomaly and $\,\omega\,$ is the argument of the pericentre; so the sum $\,f+\omega\,$ renders the angle between the line connecting the bodies and the ascending node. The rotation angle of the body, $\,\theta\,$, is reckoned from the same fiducial direction as the node $\,\Omega\,$; so the difference $\,\theta-\Omega\,$ is the angle between the minimal-inertia axis $\,x\,$ and the line of nodes. For a finite inclination (obliquity) $\,i\,$, the angles $\,f+\omega\,$ and $\,\theta-\Omega\,$ belong to different planes. However, in the limit of vanishing $\,i\,$, they are in the same plane, and their difference
 $\,\psi=(f+\omega)-(\theta-\Omega)\,$ gives the separation between the minimal-inertia axis $\,x\,$ and the direction towards the perturber.\vspace{3mm}}
 \label{Figure}}
 \end{figure}

 In Figure \ref{Figure}, the rotation angle $\,\theta\,$ of an extended body is reckoned from the line of apsides to the largest-elongation axis $\,x\,$, the one corresponding to the minimal moment of inertia $\,A\,$. The perturber exerts on the extended body a torque $\,\vec{\mathcal{T}}^{^{(TRI)}}$  due to the body's permanent triaxiality,
 and a torque $\,\vec{\mathcal{T}}^{^{(TIDE)}}\,$ due to tides in it. Rotation and longitudinal libration of the extended body about its major-inertia axis $\,z\,$ are then defined by these torques' polar components:
  \ba
 \stackrel{\bf\centerdot\,\centerdot}{\theta~}\,=~\frac{\,{\cal{T}}^{\rm{^{\,(TRI)}}}_{polar}\,+~{\cal{T}}^{\rm{^{\,(TIDE)}}}_{polar}}{C~~}~\,.
 \label{eq.eq}
 \label{201}
 \label{eq:despinning}
 \ea
  %
 The spin history is described by equation (\ref{eq.eq}) with initial conditions on $\,\theta\,$ and $\,\stackrel{\bf\centerdot}{\theta~}\,$. It is also influenced by  disruptive events. In our case, we assume that the secondary has already been caught into a spin-orbit resonance.
 , and that the transient (free) librations have already been damped.
 So we do not expect the initial conditions to play a significant role.

 Both the permanent triaxiality and the tidal distortion can be visualised as a ``bulge'', either stationary or moving across the volume of the planet. For many bodies, their permanent triaxiality can be approximated by an ellipsoid, i.e., by one double bulge. In distinction from this, the tidal bulge is a system of superimposed ripples of various shapes, running around the body at different angular rates. When a bulge is lagging behind the central line connecting the two bodies, the resulting component of the torque is accelerating, or spinning up. On the contrary, a bulge leading the central line in the direction of orbital motion is decelerating, or spinning down (Murray \& Dermott 1999).

 Each principal moment of inertia, generally, contains an oscillating component caused by the tides. These components can be neglected (and the moments of inertia $\,A,\,B,\,C\,$ can be set constant) in the situations where the tidal torque is much smaller than the permanent torque. This and other approximations adopted in our treatment are explained in Section \ref{appro} below.

 \subsection{The torque generated by the permanent triaxiality\label{2.2}}

 \subsubsection{Basic formulae}

 The torque due to the permanent triaxiality is approximated with its quadrupole ($\,\sim r^{\,-3\,}\,$) part (Danby 1962):
 \bs
 \ba
 {\cal{T}}^{\rm{^{\,(TRI)}}}_{polar}~=~\frac{3}{2}~(B-A)~\frac{{G}\,M^{\,*}}{r^3}~\sin2\psi~\,~,
 \label{202a}
 \ea
 where $\,G\,$ is the Newton gravity constant,  $\,M^{\,*}\,$ is the perturber's mass, $\,r\,$ is the distance between the centres of mass, while $\,\psi\,$ is the angle between the minimal-inertia axis $\,x\,$ of the extended body and the direction towards the perturber (see Figure \ref{Figure}).

 When the inclination (obliquity) $\,i\,$ is small, $\,\psi\,$ can be approximated with a sum of the angle between the ascending node and the $\,x$-axis and the angle between the node and the direction to the perturber: $\,\psi\,\approx\,(f\,+\,\omega)\,+\,(\Omega\,-\,\theta)\,$. In this approximation,
 \ba
 {\cal{T}}^{\rm{^{\,(TRI)}}}_{polar}~
 =
 ~\frac{3}{2}~(B-A)~\frac{G\,M^*}{r^3}~\sin2(f+\omega+\Omega-\theta)\,~.
 \label{tri.eq}
 \label{202b}
 \ea
 \label{202}
 \es
 For completeness, in Appendix D we present a full development of the polar torque in terms of the orbital elements, for an arbitrary inclination and in all multipoles. For now, however, the approximation (\ref{202b}) is sufficient.
 \begin{deluxetable}{lr}
 \tablecaption{Symbol key \label{Table}}
 \tablewidth{0pt}
 \tablehead{
 \multicolumn{1}{c}{Notation}  &
 \multicolumn{1}{c}{Description}\\
 }
 \startdata
  $R$ & \dotfill the mean radius of the tidally perturbed body \\
 $C$ & \dotfill the maximal moment of inertia of the tidally perturbed body\\
 $B$ & \dotfill the middle moment of inertia of the tidally perturbed body\\
 $A$ & \dotfill the minimal moment of inertia of the tidally perturbed body\\
 ${\cal{T}}^{\rm{^{\,(TIDE)}}}_{polar}$ & \dotfill the polar component of the tidal torque acting on the perturbed body\\
 ${\cal{T}}^{\rm{^{\,(TRI)}}}_{polar}$ & \dotfill $\,.\,.\,.\,$ the polar component of the torque due to the triaxiality\\
 $M$ & \dotfill the mass of the tidally perturbed body \\
 $M^*$ & \dotfill the mass of the perturber \\
 $r$ & \dotfill the instantaneous distance between the two bodies \\
 $a$ & \dotfill the semimajor axis of the perturber\\
 $f$ & \dotfill the true anomaly  of the perturber \\
 $e$ & \dotfill the orbital eccentricity  of the perturber \\
 ${\cal{M}}$ & \dotfill the mean anomaly  of the perturber \\
  $n\equiv\,\stackrel{\bf\centerdot}{\cal{M}\,}$ & \dotfill the anomalistic mean motion ~\\
 $\theta$ & \dotfill the rotation angle of the tidally perturbed body\\
 $\stackrel{\bf{\centerdot}}{\theta\,}$ & \dotfill the rotation rate of the tidally perturbed body\\
  $\gamma$ & \dotfill the libration angle of the extended body\\
 ${\cal{A}}$ & \dotfill the magnitude of libration of the extended body\\
 $z$ & \dotfill the number of a spin-orbit resonance ~\\
 $G$ & \dotfill Newton's gravitational constant \\
 $J_s$ & \dotfill the Bessel functions of the first kind\\
 $\chi$ & \dotfill the principal frequency of small-magnitude free libration\\
 \enddata
 \end{deluxetable}

 \subsubsection{Qualitative considerations\label{restorative}}

 As demonstrated below in Section \ref{libra}, expression (\ref{202b}) for the quadrupole part of the polar torque is equivalent to
 \ba
 \mathcal{T}_{polar}^{^{(TRI)}}=~
 \frac{3}{2}~(B-A)~\frac{G\,M^*}{a^3}\sum_{q=-\infty}^{+\infty}~G_{20q}(e)~\sin\left(\,2\,\left[
 \left(1\,+\,\frac{q}{2}\right)\,{\cal{M}}\,+\,\omega\,+\,\Omega\,-\,\theta\,\right]\,\right)\quad,\quad
 \label{this}
 \ea
 with the eccentricity functions linked to the Hansen coefficients through $\,G_{lpq}(e)\,=$
 $X_{l-2p+q}^{\,-(l+1),~l-2p}(e)\,_{\textstyle{_{\textstyle{.}}}}$
 The expression (\ref{this}) is a truncated version of a complete expansion (\ref{survivor}) provided in Appendix D.

 To get a touch of this theory, consider a case where (a) the inclination is negligible and the nodal precession is slow, and (b) the apsidal precession is slow. Then, for simplicity, both
 $\,\Omega\,$ and $\,\omega\,$ can be set zero, and the above expression becomes (Goldreich \& Peale 1968, eqn 14; Noyelles et al. 2014, eqn 10):
 \ba
 \mathcal{T}_{polar}^{^{(TRI)}}~=~
 \frac{3}{2}~(B-A)~\frac{G\,M^*}{a^3}\sum_{q=-\infty}^{+\infty}~G_{20q}(e)~\sin\left(\,2\,\left[\,
 \left(1\,+\,\frac{q}{2}\right)\,\cal{M}\,-\,\theta\right]\,\right)\quad.\qquad
 \label{203a}
 \label{eq:tri4}
 \label{kau.eq}
 \label{203b}
 \label{203}
 \ea


 All the terms of $\,\mathcal{T}_{polar}^{^{(TRI)}}$ are periodic (including the resonant term which is zero).
 A term number $\,q\,$ drops out when either $\,2\,\theta\simeq (2+q)\,{\cal M}\,$~or $\,2\,\theta\simeq (2+q)\,{\cal M}\,+\,\pi\,$, both situations corresponding to the same spin-orbit resonance. In the first case, the longest dimension is pointing at the perturber as it is passing the pericentre. In the second case, the extended body is traversing the pericentre, being sidewards to the perturber.

\begin{figure}[h]
 \begin{minipage}{135mm}
 {\hspace{2.0cm}}
 \includegraphics[width=135mm]{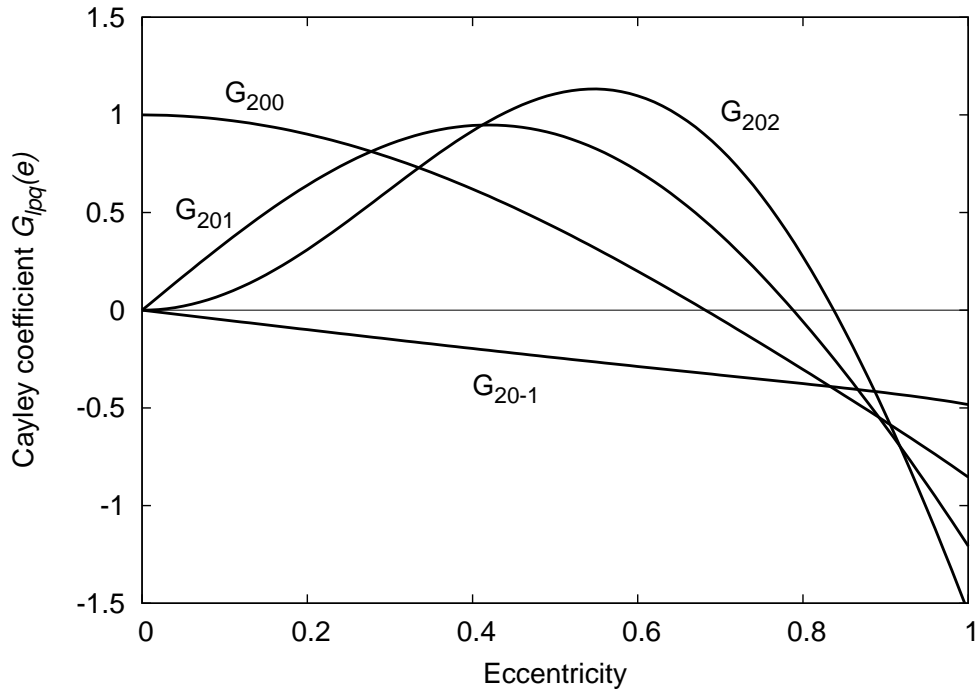}\vspace{17mm}\\
  \end{minipage}
 \hspace{2pc}
 \caption{The eccentricity functions $\,G_{20q}(e)\,$, for $\,q = -1,\,0,\,1,\,2\,$.
 ~~For $\,q=\,-\,2\,$, the function is identically zero: $\,G_{20,-2}(e)\,=\,0\,$. }
 \label{G.fig}
 \end{figure}

 Observe that the amplitudes of the periodic terms in series (\ref{203}) differ by the coefficients $\,G_{20q}(e)\,$ only. The eccentricity functions with \footnote{~Recall that for $\,q\,=\,-\,2\,$ the function is identically zero: $\,G_{20,-2}(e)\,=\,0\,$. Aside from such occasional exceptions, the eccentricity functions satisfy the rule $\,G_{lpq}(e)\,\propto\,e^{|q|}\,$.}
 $\,q=-1,0,1,2\,$ are shown in Figure \ref{G.fig}. For most of the tidally interacting pairs in the Solar system, the eccentricity is small ($\,e<0.1\,$) and the term with $\,q=0\,$ is by far the largest.$\,$\footnote{~Special is the case of Mercury because of its higher orbital eccentricity $\,e=0.20563\,$ and also because of this planet being trapped in the 3:2 resonance ($q=1$). So, for Mercury, the terms with $\,q\,=\,-\,4\,,\,.\,.\,.~\,,\,6\,$ should be taken into account, see Noyelles et al. (2014).} Once a planet is caught into a resonant state with $\,2\,\theta\approx (2\,+\,q)\,{\cal M}$, the sine function in expression (\ref{kau.eq}) can be approximated with its argument. Then, provided the value of $\,G_{20q}(e)\,$ is positive, the sign of $\;\partial \mathcal{T}_{polar}^{^{(TRI)}}/\partial \theta\;$ is negative, so the torque becomes restorative~---~which is a necessary condition for a stable resonance. Another interesting possibility is that of the resonant state with $\,2\,\theta\approx (2\,+\,q)\,{\cal M}+\pi\,$. In that case, the torque is restorative for $\,G_{20q}(e)<0\;$ (Goldreich \& Peale 1968, Makarov 2012).

 \subsubsection{More about the approximations made\label{appro}}

 Aside from the small-obliquity assumption, two more caveats are in order here.\vspace{2mm}

 1.~ The expressions (\ref{202}) for the torque $\,\mathcal{T}_{polar}^{^{(TRI)}}\,$ include a quadrupole part only. Usually acceptable for planets and planetary moons, this approximation is not necessarily good for close-in asteroidal binaries (Taylor \& Margot 2010).\vspace{2mm}


 2.~ To evaluate the libration in the angular velocity, $\,\stackrel{\bf{\centerdot}}{\theta\,}$, one has to insert the torques into equation (\ref{201}) and to integrate it once. Since we are interested in the case where
 \ba
 |\,{\cal{T}}^{\rm{^{\,(TIDE)}}}_{polar}\,|\,\ll\;|\,{\cal{T}}^{\rm{^{\,(TRI)}}}_{polar}\,|\;\;,
 \label{ineq}
 \ea
 $\,{\cal{T}}^{\rm{^{\,(TIDE)}}}_{polar}\,$ can be neglected in (\ref{201}) when we calculate the main librations, those due to triaxiality. For the same reason, we may set the
 matrix of inertia constant in time. E.g., while the maximal moment of inertia $\,C\,$ contains an oscillating component caused by the tides, our condition (\ref{ineq}) permits us to assume that it is much smaller than the permanent part of $\,C\,$.

 \subsection{Libration in longitude\label{libra}}

 An approximate description of longitudinal libration inside spin-orbit resonances is usually derived by neglecting the tidal torque and reducing the ensuing equation $\;C\,\ddot{\theta}= {\cal{T}}^{\rm{^{\,(TRI)}}}_{polar}\,$ to a simple forced harmonic oscillator.

 \subsubsection{From the true anomaly to the mean anomaly}

 Using expression (\ref{202}), we write the equation of motion as
 \bs
 \ba
 \ddot{\theta} &=& \omega_0^2\;\frac{a^3}{r^3}\;\sin \left[\,2\,f-\,2\,(\theta\,-\,\Omega\,-\,\omega)\,\right]
 \label{}\\
               &=& \omega_0^2\;\frac{a^3}{r^3}\;\left[\sin 2f\;\cos 2(\theta\,-\,\Omega\,-\,\omega)\,-\,\cos 2f\;\sin 2(\theta\,-\,\Omega\,-\,\omega) \right]\;\,,
 \label{tisserand}
 \ea
 \es
 where
 \begin{equation}
 \omega_0^2\;=\;\frac{3}{2}~\frac{B-A}{C}~\frac{G\;M^*}{a^3}\;\,.
 \end{equation}
 At this point, we need the following Fourier series (e.g., Duriez 2002):
 \begin{eqnarray}
 \left(\frac{a}{r}\right)^3\;\sin(2f) & = & \sum_{k=-\infty}^{+\infty}\,X_k^{-3,\,2}(e)\;\sin(k\cal{M})\,~,
 \label{eq:duriezsin}
 ~\\
 \left(\frac{a}{r}\right)^3\;\cos(2f) & = & \sum_{k=-\infty}^{+\infty}\,X_k^{-3,\,2}(e)\;\cos(k\cal{M})\,~,
 \label{eq:duriezcos}
 \end{eqnarray}
 with $\,X_{k}^{-3,2}(e)\;=\;G_{20(k-2)}(e)\,$.
 Insertion of these series into equation (\ref{tisserand}) results in
 \ba
 \ddot{\theta}
             \;=\;\omega_0^2\,\sum_{k=-\infty}^{+\infty}\,
              G_{20(k-2)}(e)
              \;\sin \left[\,k\,{\cal{M}}\,-\,2\,(\theta\,-\,\Omega\,-\,\omega)\,\right]\,\;.
 \label{landau}
 \ea

 \subsubsection{Definition of a spin-orbit resonance\label{small}}

 To discuss small longitudinal libration near a spin-orbit resonance, we first need to agree on what a resonance is. By definition, an $\,lmpq\,$ spin-orbit resonance is a situation where one or several terms in the expansion for a torque go through zero. We say $\,$``{{$\,${\underline{a}} $\,$torque$\,$}}''$\,$,
 $\,$because one and the same sufficient condition,$\,$\footnote{~The
  rotation angle $\,\theta\,$ and the node $\,\Omega\,$ must be reckoned from the same fiducial direction (usually, the vernal equinox), see Footnote \ref{important} in Appendix D.
  }
 \ba
 (l-2p)\,\omega\,+\, (l-2p+q){\cal M}\,+\,m\,\Omega\,-\,m\,\theta\,=\,-\,N\,\pi\;\,,
 \label{above}
 \ea
 with an integer $\,N\,$, ensures a resonance for both the tidal torque and for the principal (ellipsoidal) part of the torque caused by the permanent triaxiality.

  As explained in Appendix D, an $\,lmpq\,$ term of the Fourier expansion (\ref{ellipsoid}) of the permanent-triaxiality-caused torque acting on a triaxial ellipsoid contains the sine of the linear combination standing on the left-hand side of (\ref{above}). For bodies strongly deviating from a triaxial ellipsoid, the situation will be more complex because of additional terms showing up in the expression for the torque. We shall not go there, because realistic planets and moons are not very different in shape from triaxial ellipsoids, so the said additional terms are relatively small and their role is limited.

  Also, as demonstrated in Appendix B3, an $\,lmpq\,$ term of the Fourier expansion (\ref{torque}) of the tidal torque contains the sine of the phase lag $\,\epsilon_l\,=\,\epsilon_l(\omega_{lmpq})\,=\,\omega_{lmpq}\,\Delta t_l(\omega_{lmpq})\,$, where $\,\Delta t_l(\omega_{lmpq})\,$ is the appropriate time lag and $\,\omega_{lmpq}\,$ is the appropriate tidal Fourier mode. This tells us that an $\,lmpq\,$ spin-orbit resonance corresponds to the phase lag transcending zero, i.e.,
  to $\,0\,=\,\omega_{lmpq}\,=\,(l-2p)\,\stackrel{\bf\centerdot}{\omega\,}+\,(l-2p+q)\stackrel{\bf\centerdot}{\cal M\,}+\,m\,\stackrel{\bf\centerdot}{\Omega\,}-\,m\,\stackrel{\bf\centerdot}{\theta\,}$. Up to an irrelevant constant, this may be written as (\ref{above}). This way, equality (\ref{above}) can serve as a condition of resonance for both torques.


 All in all, we say that a body is crossing an $\,lmpq\,$ spin-orbit resonance when its rotation angle is transcending the value of \footnote{~Here ``crossing a resonance'' means: despinning (or spinning up) through a resonance. Usually, the resonance condition is written in terms of angular rates: $\,\stackrel{\bf\centerdot}{\theta\,}=\,\frac{\textstyle l-2p+q}{\textstyle m}\,\stackrel{\bf\centerdot}{\cal M\,}+\,\frac{\textstyle l-2p}{\textstyle m}\,\stackrel{\bf\centerdot}{\omega\,}+\,\stackrel{\bf\centerdot}{\Omega\,}\,$. While the so-written condition would be sufficient for our purposes, we prefer to express it through the angles. This is done to emphasise that it is not only the rate $\,\stackrel{\bf\centerdot}{\theta\,}$ but also the value of the rotation angle $\,\theta\,$ in the pericentre, that defines the fate of a rotator crossing a resonance~---~i.e., whether the rotator gets captured into the resonance or transcends it (Makarov 2012).
 }
 \ba
 \theta_{res}\;=\;\frac{l-2p+q}{m}\,{\cal M}\,+\,\frac{l-2p}{m}\,\omega\,+\,\Omega\,+\,\frac{N}{m}\,\pi\,\;.
 \label{res}
 \ea
 When a triaxial rotator gets trapped in this resonance and is librating in it, its rotation angle $\,{\theta}\,$ becomes a sum of the now-constant part $\,\theta_{res}\,$ and a longitudinal libration angle $\,\gamma\;$:
 \ba
 \theta\;=\;\theta_{res}\;+\;\gamma\;=\;\frac{l-2p+q}{m}\,{\cal M}\,+\,\frac{l-2p}{m}\,\omega\,+\,\Omega\,+\,\frac{N}{m}\,\pi\,+\,\gamma\,\;.
 \label{defff}
 \ea
 The term $\,\gamma\,$ comprises physical libration in longitude and a constant bias.$\,$\footnote{~In lunar studies, a common convention is to write the resonant rotation angle as
 $$
  {\theta}\,=~180^o~+~\Omega~+~{\omega}~+~{\cal{M}}~+~{{\gamma}}\,~,
 $$
 with extra $\,180^o\,$ added. This convention is employed, e.g., in Eckhardt (1981) where the sum $\,\phi+\psi\,$ is the same as our $\,\theta\,$; while $\,\tau\,$ is the same as our $\,\gamma$, the libration in longitude. The largest (annual) periodic term in longitude libration is $\,91''\,$, see the Electronic Supplementary Materials to  Rambaux \& Williams (2011).

 The apparent difference between Eckhardt's convention and the convention (\ref{22218}) stems from the fact that Eckhardt's mean longitude
 $\,L\,=\,\Omega\,+\,{\omega}\,+\,{\cal{M}}\,$ stands for the mean longitude {\it{of the Moon centre as seen from the Earth's centre}}. Accordingly,
 $\,180^o+L=180^o+\Omega\,+\,{\omega}\,+\,{\cal{M}}\,$ is the orbital mean longitude {\it{of the Earth centre as seen from the Moon's centre}}.

 In Eckardt (1981), the Cartesian selenographic coordinates coincide with the lunar principal moments of inertia in the absence of elastic deformation, one of the principal axes pointing (approximately) towards the Earth. Eckhardt's addition of the $\,180^o\,$ to $\,L\,$ puts the zero longitude on the average centre of the lunar face toward the Earth. Also be mindful that Eckhardt employed mean (slightly nonosculating) orbital elements.

 Sometimes the so-chosen coordinate axes are termed as the {\it{mean Earth}} or {\it{mean rotatation coordinates}}. It should however be noted that the mean principal axes are biased $\,68''\,$ in longitude and $\,79''\,$ in latitude from the actual mean Earth direction (Rambaux \& Williams 2011), so $\,\gamma\,$ includes a small constant term.
 The bias is due to the $\,l>2\,$ degree terms of the lunar gravitational field, and is model-dependent (Williams et al. 2013).
  \label{lunar}}
  Given the extreme smallness of the bias, we shall ignore it and shall identify $\,\gamma\,$ with physical libration solely. Be mindful that $\,\gamma\,$ does $\,${\it{not}}$\,$
  include the optical libration which is already taken care of by the standard Kaula formalism.

 At low inclinations, only the terms with $\,p=0\,$ are to be kept:
 \ba
 {\theta}\,=~z\,{\cal{M}}\;+\;z\,'\,{\omega}\,+\,\Omega\,+\,\frac{N}{m}\,\pi\,+~{{\gamma}}\,~,
 \label{2218}
 \label{6}
 \ea
 \ba
 z\;=\;\frac{l\,+\,q}{m}\quad,\quad z\,'\,=\;\frac{l}{m}\;\;.
 \nonumber
 \ea
 If, above that, only the quadrupole parts of the torques matter ($\,l=m=2\,$), the rotation angle in a spin-orbit resonance becomes simply
 \ba
 {\theta}\,=\,z\,{\cal{M}}\,+\,{\omega}\,+\,\Omega\,+\,\frac{N}{2}\,\pi\,+\,{{\gamma}}\,~,
 \label{22218}
 \ea
 with $\,z\,$ now being not just rational but, importantly, semi-integer:
 \ba
 z\;=\;1\;+\;\frac{q}{2}\;\,.
 \nonumber
 \ea

 Usually, the condition of resonance is written down in terms of the angles' rates:
 \ba
 \dot{\theta}\,=\,z\,\dot{\cal{M}}\,+\,\dot{\omega}\,+\,\dot\Omega\,+\,\dot{\gamma}\,~.
 \nonumber
 \ea
 We however have written this condition also via the angles, equation (\ref{22218}), in order to emphasise that libration may, in principle, be taking place about various values of the angle $\,\theta\,$. For example, at high eccentricities, a rotation regime is available wherein the body traverses the pericentre being sidewards to the perturber, see Section \ref{restorative} above. (Here $\,${\it{available}}$\,$ implies that the triaxial torque is restoring~---~which does not by itself preclude emergence of chaos.)

 \subsubsection{Small longitudinal libration near a spin-orbit resonance}

 Insertion of expression (\ref{22218}) for the resonant $\,\theta\,$ into equation (\ref{landau}) leads us to
 \ba
 \ddot{\gamma}\;=\;\omega_0^2\;\sum_{k=-\infty}^{\infty}
 G_{20(k-2)}(e)
 \;\left[\;\sin\left(\,(k-2z)\,{\cal{M}}\,\right)\;\cos 2 \gamma
 -\;\cos\left(\,(k-2z)\,{\cal{M}}\,\right)\;\sin 2 \gamma\;\right]\,\;.
 \label{fifteen}
 \ea
 For small libration in resonance, we have
 \ba
 \cos(2 \gamma) \simeq 1\quad \mbox{and} \quad\sin(2 \gamma) \simeq 2 \gamma\,\;.
 \label{condi}
 \ea
 Under this assumption,
 and also in neglect of non-linear terms, we end up with the expression
 \begin{equation}
 \ddot{\gamma}\,+\,2\,\omega_0^2\,
 G_{20(2z-2)}(e)
 \,\gamma\;=\;\omega_0^2\,\sum_{k=-\infty}^{\infty}
 G_{20(k-2)}(e)
 \;\sin\left(\,(k-2z)\,{\cal{M}}\,\right)\,\;.
 \nonumber
 \end{equation}
 For $\,{\cal{M}} = nt\,$, this becomes the equation for a forced harmonic oscillator. Its solution is
 \begin{equation}
 \gamma(t)\;=\;{\cal{A}}\;\sin(\chi t + \phi) + \omega_0^2 \sum_{k=-\infty}^{\infty} \frac{
 G_{20(k-2)}(e)
 }{\,\chi^2\,-\,(k-2z)^2\,n^2\,}\;\sin\left(\,(k-2z)\,n\,t\,\right)\,\;,
 \label{eq1}
 \end{equation}
 with the natural frequency $\,\chi\,$ found from
 \begin{equation}
 \chi^2\;=\;2\;\omega_0^2\;
 G_{20(2z-2)}(e)
 \;=\;3~\frac{B-A}{C}~\frac{G\;M^*}{a^3}\;G_{20(2z-2)}(e)
 \,\;,
 \label{publish}
 \end{equation}
 and $\,n=\,\stackrel{\bf\centerdot}{\cal{M\,}}\,$ being the anomalistic mean motion, not the osculating one.$\,$\footnote{~Recall that the anomalistic mean motion $\,n(t)\,\equiv\,{\bf{\dot{\cal{M}}}}\,=\,{\bf{\dot{\cal{M}}}}_0(t)\,+\,n_{\textstyle{_{Kepler}}}(t)\,$ is but an approximation to the Keplerian mean motion $\,n_{\textstyle{_{Kepler}}}(t)\,\equiv\,\sqrt{G(M^*+M)\,a^{-3}(t)\,}\,$. $\,$This is explained in more detail in Appendix B to Efroimsky \& Makarov (2014).
 Also be mindful that in our derivation of equation (\ref{eq1}) it was legitimate to use the approximation $\,{\cal{M}} = nt\,$ insofar as the rate $\,{\bf{\dot{\cal{M}}}}\,=\,{\bf{\dot{\cal{M}}}}_0(t)\,+\,n_{\textstyle{_{Kepler}}}(t)\,$
 can be assumed constant over a period of libration.}

 The constants $\,{\cal{A}}\,$ and $\,\phi\,$ parameterise the free libration and are determined by the initial conditions. With a tidal torque included into the equation of motion (\ref{eq.eq}), the amplitude $\,{\cal{A}}\,$ is no longer constant but is being damped exponentially.$\,$\footnote{~The situation becomes more complicated when the tidal torque is nonlinear in frequency. In this situation, limit cycles can emerge, resulting in persistent nonvanishing free libration with a very small magnitude (Frouard \& Makarov 2018).\label{persistent}} Unless stated otherwise, we shall assume that this free libration is entirely damped, and shall be concerned with the forced libration only. The expression (\ref{eq1}) can then be cast as a sum over positive integers:
 \ba
 \gamma(t)\;=\;\sum_{j=1}^{\infty}{\cal{A}}_j\;\sin(j n t)\,\;.
 \label{}
 \ea
 In this notation, the magnitudes read as
 \bs
 \ba
 {\cal{A}}_j & = & \omega_0^2\;\frac{G_{20(j+2z-2)}(e) - G_{20(-j+2z-2)}(e)}{\chi^2\,-\,j^2\,n^2}
 \label{}\\
 \nonumber\\
       & \approx & -\;\omega_0^2\;\frac{G_{20(j+2z-2)}(e) - G_{20(-j+2z-2)}(e)}{j^2\,n^2}\,\;,
 \label{eq2}
 \ea
 \es
 where we assumed that $\,\chi^2 \ll j^2\,n^2\,$.

 \subsubsection{Two examples}

In the synchronous resonance ($z=1$), we have
\begin{equation}
\begin{split}
^{(1:1)}\gamma({\cal{M}}) \simeq  \frac{\omega_0^2}{n^2} \bigg[
\bigg( -4 e  + \frac{31}{4} e^3 + \mathcal{O}(e^5) \bigg) \sin {\cal{M}}
+ \bigg( - \frac{17}{8} e^2  + \mathcal{O}(e^4) \bigg) \sin 2{\cal{M}} \\
+ \bigg( - \frac{211}{108} e^3 + \mathcal{O}(e^5) \bigg) \sin 3{\cal{M}}
+ \mathcal{O}(e^4) \bigg]\,\;.\;\;\;
\label{negative}
\end{split}
\end{equation}
For the 3:2 spin-orbit resonance ($z=3/2$), we have
\begin{equation}
\begin{split}
^{(3:2)}\gamma({\cal{M}}) \simeq  \frac{\omega_0^2}{n^2} \bigg[
\bigg( 1 - 11 e^2  + \mathcal{O}(e^4) \bigg) \sin {\cal{M}}
+ \bigg( - \frac{1}{8} e  - \frac{421}{96} e^3 + \mathcal{O}(e^5) \bigg) \sin 2{\cal{M}} \\
+ \bigg( \mathcal{O}(e^4) \bigg) \sin 3{\cal{M}}
+ \bigg(  \frac{1}{768} e^3   + \mathcal{O}(e^5) \bigg) \sin 4{\cal{M}}
+ \mathcal{O}(e^4) \bigg]\,\;.
\end{split}
\end{equation}
  The anomalistic mean motion $\,n\,\equiv\,\stackrel{\bf{\centerdot}}{\cal{M}\,}\,$ often can be approximated with its osculating counterpart $\,\sqrt{G(M^*+M)/a^3\,}\,$, in which cases the common factor can be written down as
 \ba
 \frac{\omega_0^2}{n^2}\;=\;\frac{3}{2}~\frac{B-A}{C}~\frac{M^*}{M^*+\,M}\;\,.
 \label{iii}
 \ea

 \section{The Darwin-Kaula expansion of bodily tides}

 At large, a description of linear bodily tides consists of two consecutive steps.

 The first one is to Fourier-expand both the external tide-raising potential and the induced additional tidal potential of the distorted body. This work was begun by Darwin (1879) who wrote down several leading terms of the Fourier series for these potentials, and who also derived an expression for the quadrupole $\,${\it{dynamical}}$\,$ Love number of an incompressible homogeneous sphere.$\,$\footnote{~The formalism of $\,${\it{static}}$\,$ Love numbers was pioneered by Lord Kelvin (Thomson 1863, Thomson \& Tait 1867) for an incompressible homogeneous spherical body. In his works, Love extended this formalism to the case with compressibility (Love 1911, Chapter VIII, Eqn 16).} Almost a century later, an impressive mathematical effort undertaken by Kaula (1961) resulted in his developing a complete series (Kaula 1964).
  In the present-day notation, Darwin's work is explained by Ferraz-Mello et al. (2008). A comprehensive explanation of Kaula's development is given in
  Efroimsky \& Makarov (2013).

 The second step in the description of tides is to link each spectral component of the reaction to an appropriate component of the action. This means: (1) to determine the phase lag between a Fourier harmonic of the additional tidal potential of the distorted body and the corresponding Fourier harmonic of
 the perturbing potential, and (2) to find the ratio of the magnitudes of these two harmonics (the dynamical Love number). The frequency-dependencies of the phase lag and of the dynamical Love number are dictated by the interplay of the body's rheology and self-gravitation (Efroimsky 2012$\,$a,b, $\,$Efroimsky 2015).

 \subsection{The asterisk convention}\label{asterisk}

 The central result of the Darwin-Kaula theory of bodily tides is constituted by the Fourier expansions of the tide-raising potential $\,W\,$ and of the ensuing additional tidal potential $\,U\,$ of a perturbed near-spherical body. Both these potentials are expressed through vectors $\,{\erbold}^{\;*}\,$ and $\,\Rbold\,$ pointing from the perturbed body's centre. Vector $\,{\erbold}^{\;*}\,$ is the position of the external perturber on its orbit as seen from the perturbed body, while the vector $\,\Rbold = (R,\phi,\lambda)\,$ singles out a point on the perturbed body's surface.

 A convention established by Kaula (1964) prescribes to denote the coordinates of the perturber by letters with asterisk, reserving variables with no asterisk to a test particle.$\,$\footnote{~In the afore cited work by Ferraz-Mello et al. (2008), an opposite convention is used. There, orbital elements with asterisk correspond to a test particle, while those without asterisk correspond to the perturber.} In a particular case, the test body can coincide with the perturber, in which
 situation the quantities with asterisk are set to coincide with their counterparts with no asterisk. This, however, is done in the end of Kaula's development, while the
 development itself is performed in understanding that the perturber and the test body are, generally, two different objects.

 In our calculations, the mass of the perturber will be denoted with $\,M^{\,*}\,$ and its position with $\,\erbold^{\;*}\,$. When the perturber and the test particle are considered in our calculations as two separate bodies, we shall follow Kaula's convention, denoting the perturber's coordinates by letters with asterisk: $\,\erbold^{\;*}\,=$ $(r^{\;*},\,\lambda^{\;*},\,\phi^{\;*})=$ $(\,a^{\;*},\,e^{\;*},\,i^{\;*},\,\Omega^{\;*},\,\omega^{\;*},\,{\cal M}^{\;*}\,)\,$.

 Since in our treatment the test body and the perturber are $\,${\it{ab initio}}$\,$ the same body, we shall write its position as: $\,\erbold^{\;*}\,=\,(\,a,\,e,\,\inc,\,\Omega,\,\omega,\,{\cal M}\,)\,$, leaving the orbital elements without asterisk.
 However, in the Appendices B and C, when deriving the necessary formulae, we keep the distinction using the asterisk convention.


 \subsection{The tide-raising potential}

 As explained in Appendix B.2, equation (\ref{sum}), the potential generated by a perturber in a surface point $\,\Rbold = (R,\phi,\lambda)\,$ of an extended body can be written as
 \ba
 \nonumber
  W(\eRbold\,,~\erbold^{~*})\;=~-~\frac{G\,M^*}{a}~\sum_{l=2}^{\infty}~\left(\,\frac{R}{a}\,\right)^{\textstyle{^l}}\sum_{m=0}^{l}~\frac{(l - m)!}{(l + m)!}~
  \left(\,2
  \right. ~~~~~~~~~~~~~~~~~~~~~~~~~~~~~~~~~~~~~~\\  \nonumber\\  \left.~~~
   -\;\delta_{0m}\,\right)\;P_{{\it{l}}m}(\sin\phi)\;\sum_{p=0}^{l}F_{lmp}(\inc)\;\sum_{q=\,-\,\infty}^{\infty}G_{lpq}(e)
 ~\cos\left(v_{lmpq}-m(\lambda+\theta)+\psi_{lm}\right)~~~,~\qquad\quad
 \label{1b}
 \label{213b}
  \label{213}
 \label{1}
 \ea
 where supplementary phases are given by
 \ba
 \psi_{lm}~=~\left[\,(\,-\,1)^{\,l-m}\,-\,1\,\right]~\frac{\pi}{4}\;~~\qquad~ ~\qquad~ ~\qquad~
 \label{2}
 \label{214}
 \ea
 and the auxiliary quantities $\,v_{lmpq}\,$ are defined as
 \ba
 v_{lmpq}\;\equiv\;({\it l}-2p)\,\omega\,+\, ({\it l}-2p+q){\cal M}\,+\,m\,\Omega~~~,
 \label{3}
 \label{215}
 \ea
 with $\,G
 \,$ being Newton's gravity constant, $\,G_{lpq}(e)$ and $\,F_{lmp}(\inc)\,$ being the eccentricity and inclination functions, and
 $\,P_{lm}(\sin\phi)\,$ being the associated Legendre polynomials.

 Insertion of (\ref{3}) into (\ref{1}) demonstrates that the tidal potential depends upon the difference $\,\Omega\,-\,\theta\,$ between the longitude of the node and the rotation angle. Neither of these two angles shows up separately from another, because both are reckoned from a common reference direction fixed in the equatorial plane.$\,$\footnote{~The reference direction is coprecessing with the equator but is not corotating with it.} A change in the reference direction is a
 purely mathematical convention, which should have no influence on the description of tides. This invariance is reflected by these angles entering the tidal theory as a difference $\,\Omega-\theta\,$ only. For a vanishing inclination $\,i\,$, the argument of the pericentre obviously \footnote{~Obvious on general grounds, this fact is not immediately apparent in the maths, because in expression (\ref{3}) the pericentre is multiplied by $\,(l-2p)\,$, while the node is accompanied with a factor of $\,m\,$. It however can be proven (Gooding \& Wagner 2008, Section 9.2) that $\,F_{lmp}(0)=0\,$ unless $\,l-2p=m\,$. Thence, for a zero $\,i\,$, only the terms with $\,l-2p=m\,$ survive in series (\ref{1}) and the angles indeed show up in the combination $\,\Omega+\omega-\theta\,$.} must join these two angles, so the three angles must enter the theory in the combination $\,\Omega+\omega-\theta\,$.

 \subsection{The additional tidal potential of the perturbed body}

 As demonstrated in Appendix B, the induced tidal potential of the perturbed body, caused by its tidal deformation, can be written in a manner similar to expression (\ref{1}):
 \ba
 \nonumber
 \nonumber
 U(\erbold\,,\;\erbold^{\;*})&=&
 -\;\frac{G\,M^*}{a}\;\sum_{l=2}^{\infty}\;\left(\,\frac{R}{r}\,\right)^{\textstyle{^{l+1}}}
  \left(\,\frac{R}{a}\,\right)^{\textstyle{^{
  l}}}\sum_{m=0}^{l}\;\frac{({\it l} - m)!}{({\it l} + m)!}\;
  \left(\,2~-\;\delta_{0m}\,\right)\;P_{lm}(\sin\phi)~~~~\\
  \nonumber\\
  &~&\left.~\right.\sum_{p=0}^{l}F_{lmp}(\inc)\;\sum_{q=\,-\,\infty}^{\infty}\;G_{lpq}
  (e)~k_l~
  \cos\left(\,v_{lmpq}~-\,m\,(\lambda~+~\theta)~+~\psi_{lm}~-~\epsilon_l
  \, \right)~~,~~\qquad~\,
 \label{10b}
 \label{10}
 \ea
 $k_l\,$ being the dynamical Love numbers and $\,\epsilon_l\,$ being the phase lags. The frequency-dependencies of the Love numbers and phase lags are determined by the rheological properties and self-gravitation of the tidally perturbed body.

 The Love numbers and phase lags depend on the tidal Fourier modes. As we shall see in Section \ref{interpretation}, in the absence of libration these modes are parameterised with the four indices $\,lmpq\,$. Conventionally, these modes are named as $\,\omega_{lmpq}\,$, so the Love numbers and phase lags can be written as $\,k_l(\omega_{\textstyle{_{lmpq}}})\,$
 and $\,\epsilon_l(\omega_{\textstyle{_{lmpq}}})\,$. For a homogeneous sphere, the functional form of these dependencies is defined by the degree $\,l\,$ solely,
 while the dependence on $\,m,\,p,\,q\,$ comes from the argument of these dependencies, the tidal mode $\,\omega_{\textstyle{_{lmpq}}}\,$. The functional form of the
 dependencies will be parameterised also by the order $\,m\,$, if the oblateness is taken into account: $\,k_{lm}(\omega_{\textstyle{_{lmpq}}})\,$,
 $\,\epsilon_{lm}(\omega_{\textstyle{_{lmpq}}})~$ ~---~see Dehant (1987a,b) and references therein.

 It will also be demonstrated shortly that under forced libration the Love numbers and phase lags become functions $\,k_l(\beta_{\textstyle{_{lmpqs}}})\,$
 and $\,\epsilon_l(\beta_{\textstyle{_{lmpqs}}})\,$ of different Fourier modes $\,\beta_{lmpqs}\,$. These will
 be numbered not with four but with five independent integers~---~and will incorporate the modes $\,\omega_{\textstyle{_{lmpq}}}\,$ as a subset.

 \subsection{Tidal Fourier modes and phase lags.\\
 Limitations of the Darwin-Kaula theory \label{interpretation}}

 Although Kaula himself never bothered to write down the spectrum of tidal modes, that step is easy to carry out.

 Introducing for brevity the notation
 \ba
 B_{lmpq}\,=\,v_{lmpq}\,-\,m\,(\lambda\,+\,\theta)\,+\,\psi_{lm}\,\;,
 \label{}
 \ea
 we recall that in a fixed point $\,(R,\,\phi,\,\lambda)\,$ of the surface of the perturbed body the tide-raising potential is expanded into series (\ref{1b}) whose $\,lmpq\,$ term contains a multiplier
 \ba
 \cos B_{lmpq}\;=\;\cos(v_{lmpq}\,-\,m\,(\lambda\,+\,\theta)\,+\,\psi_{lm})
 \,\;.
 \label{cos}
 \ea

 Over timescales shorter than the period of apsidal motion,$\,$\footnote{~Our caveat pertains to the apsidal motion solely, not to the nodal one. The reason for this will
 become clear shortly, when we insert the expression (\ref{3}) for $\,v_{lmpq}\,$ and the expression (\ref{res}) for the resonant angle into the expression (\ref{5}) for the quantities $\,\omega_{\textstyle{_{lmpq}}}\,$. After this insertion, the longitude of the node will drop out from the expression for $\,\omega_{\textstyle{_{lmpq}}}\,$. So the nodal evolution does not influence the interpretation of $\,\omega_{\textstyle{_{lmpq}}}\,$ in resonances.} the argument of the cosine can be linearised:
 \ba
 \nonumber
 B_{lmpq}(t)&=&B_{lmpq}(t_0)\;+\;(t\,-\,t_0)\,\dot{B}_{lmpq}\\
 \label{B}\\
 \nonumber
 &=& \left[\,v_{lmpq}(t_0)\,-\,m\,\left(\lambda\,+\,\theta^*(t_0)\,\right)\,+\,\psi_{lm}\,\right]\;+\;(t\,-\,t_0)\,\frac{d\,}{dt}\,(v_{lmpq}\,-\,m\,\theta)\quad.\quad
 \ea
 So the cosines become:
 \ba
 \cos B_{lmpq}\;=\;\cos\left(\;\left[\,v_{lmpq}(t_0)\,-\,m\,\left(\lambda\,+\,\theta(t_0)\,\right)\,+\,\psi_{lm}\,\right]\,+\,(t\,-\,t_0)\,\omega_{lmpq}\,\right)\,\;,
 \label{square}
 \ea
 where
 \ba
 \nonumber
 \omega_{lmpq}\,\equiv\,\frac{d\,}{dt}\,(v_{lmpq}\,-\,m\,\theta)&=&(l-2p)\,\dot{\omega}\,+\,(l-2p+q)\,\dot{\cal{M}}\,+\,m\,(\dot\Omega\,-\,\dot\theta)\\
 \label{5}\\
 \nonumber
                                                                    &\approx & (l-2p+q)\,\dot{\cal{M}}\,-\;m\,\dot{\theta}\,\;.
 \ea

  We have a right to treat the quantities $\,\omega_{lmpq}\,$ as tidal modes, because:\\
  ~\\ {\it
  (a) ~the expression in square brackets in the expression (\ref{square}) is secular, i.e., changes at frequencies much lower than $\,\omega_{lmpq}\,$;\\
  ~\\
  (b) ~the quantities $\,\omega_{lmpq}\,$ themselves are secular: $\;|\,\dot{\omega}_{lmpq}/\omega_{lmpq}\,|\,\ll\,|\,\omega_{lmpq}\,|\,$}\\

 The question now becomes whether the above derivation stays valid in the presence of libration, i.e., in a situation where $\,\theta\,$ contains a varying term. Generally, the
 answer to this question is negative. Indeed, when $\,\theta\,$ incorporates a ``smooth" part $\,\theta_{res}\,$ and an oscillating part $\,{\cal{A}}\,\sin{\cal{M}}\,$, an $\,lmpq\,$ term of the series (\ref{ergo}) will contain, instead of the multiplier (\ref{cos}), a multiplier looking as
 \ba
 \cos (B_{lmpq}\,-\,{\cal{A}}\,m\,\sin{\cal{M}})\;=\;\cos(v_{lmpq}\,-\,m\,(\lambda\,+\,\theta_{res})\,+\,\psi_{lm}\,-\,{\cal{A}}\,m\,\sin{\cal{M}})\,\;,
 \label{}
 \ea
 where we include into $\,B_{lmpq}\,$ only the ``smooth'' terms:
 \ba
 B_{lmpq}\,=\,v_{lmpq}\,-\,m\,(\lambda\,+\,\theta_{res})\,+\,\psi_{lm}\,\;
 \label{}
 \ea
 and keep the oscillating part separate.

 In the case of forced libration, the input $\,-\,{\cal{A}}\,m\,\sin{\cal{M}}\,$ oscillates at a rate $\,n\equiv\dot{\cal{M}}\,$ comparable to the rate $\,\omega_{lmpq}\,$ at which the ``smooth" part
 $\,v_{lmpq}\,-\,m\,\theta_{res}\,$ is changing. For this reason, the linearisation procedure employed in (\ref{B}) would no longer render quantities interpretable as Fourier modes. Indeed, had we tried to follow that procedure, we would have arrived at the ``Fourier modes''
 \ba
 \omega_{lmpq}\,=\,(l-2p)\,\dot{\omega}\,+\,(l-2p+q)\,\dot{\cal{M}}\,+\,m\,(\dot\Omega\,-\,\dot\theta)\,-\,{\cal{A}}\,m\,n\,\cos n t
 \label{quantities}
 \ea
 which are $\,${\it{not}}$\,$ evolving adiabatically, but are changing at a rate comparable to themselves. The above condition (a) is not obeyed, so the quantities (\ref{quantities}) are not secular and cannot be interpreted as Fourier modes. This points at a limitation of the Darwin and Kaula theory of tides: this theory is inapplicable under libration.

 To find the actual Fourier modes, we must resort to the relation
 \ba
 \cos(\,B_{lmpq}\,-~{\cal A}~m~\sin {\cal M}\,)~=\,\sum_{s=-\infty}^{\infty}\,J_{s}(m{\cal A})~\cos(B_{lmpq}\,-\,s\,{\cal M})\,~,
 \label{}
 \ea
 where $\,J_{s}(x)\,$ are the Bessel functions of integer order. To the best of our knowledge, this work has never been done in the literature hitherto.

 More generally, under libration at a frequency $\,\nu\,$, this expression would read as
 \ba
 \cos(\,B_{lmpq}\,-~{\cal A}~m~\sin {\nu}t\,)~=\,\sum_{s=-\infty}^{\infty}\,J_{s}(m{\cal A})~\cos(B_{lmpq}\,-\,s\,{\nu}\,t)\,~.
 \label{33}
 \ea

 \section{Generalisation of the Darwin-Kaula theory\\ to bodies librating in spin-orbit resonances\label{gen}}

 \subsection{Fourier tidal modes caused by libration in longitude}\label{4.1}

 In formulae (\ref{1b}) and (\ref{10}) for the potentials, each term is proportional to the cosine of the expression
 \ba
 v_{lmpq}~-~m~(\lambda\,+\,\theta)~+~\psi_{lm}~=~(l-2p)~\omega~+~(l-2p+q)~{\cal M}~+~m~(\Omega\,-\,\lambda\,-\,\theta)~+~\psi_{lm}\,~.\qquad
 \label{111}
 \ea
 In a commensurability parameterised by rational numbers $\,z,\,z\,'\,$, the above expression should be combined with formula (\ref{6}) for the resonant rotation angle:
 \bs
 \ba
 &&v_{lmpq}~-~m\,(\lambda\,+\,\theta)~+~\psi_{lm}\;=
 \nonumber\\
 \nonumber\\
 & &(l\,-\,2\,p\,-\,m\,z\,')\,\omega\,+\,(l\,-\,2\,p\,-\,m\,z\,+\,q)\,{\cal M}\,+\,m\,(\,-\,\lambda\,-\,\gamma)\,-\,N\,\pi\,+\,\psi_{lm}\;=\\
 \label{112a}
 \nonumber\\
 & &(l\,-\,2\,p\,-\,m\,z\,')\,\omega\,+\,(l\,-\,2\,p\,-\,m\,z\,+\,q)\,{\cal M}\,-\,m\,\lambda\,-\,N\,\pi\,+\,\psi_{lm}\,-\,{\cal A}\,m~\sin {\cal M}~~,~\qquad\qquad
 \label{112b}
 \ea
 \label{112}
 \es
 where the total libration angle has been approximated with $\,{\cal A}\,\sin {\cal M}\,$. Our next step is to separate timescales as
 \ba
 v_{lmpq}~-~m~(\lambda\,+\,\theta)~+~\psi_{lm}~=~B_{lmpq}~-~{\cal A}~m~\sin {\cal M}\,~,
 \label{113}
 \ea
 where the term $~-\,{\cal A}\,m\,\sin {\cal M}\,=\,-\,{\cal A}\,m\,\sin ({\cal M}_0\,+\,\stackrel{\bf\centerdot}{\cal M\,}t)~$ evolves much faster than the remaining sum
 \ba
 \nonumber
 B_{lmpq}&\equiv& v_{lmpq}~-~m~(\lambda\,+\,\theta_{res})~+~\psi_{lm}\\
 \label{114}\\
 &=&(l\,-\,2\,p\,-\,m\,z\,')\,\omega\,+\,(l\,-\,2\,p\,-\,m\,z\,+\,q)\,{\cal M}\,-\,m\,\lambda\,-\,N\,\pi\,+\,\psi_{lm}\,~.
 \nonumber
 \ea
 Over timescales much shorter than the period of periapse motion (be it retardation, advance or libration), the slower component may be linearised:
 \footnote{~In a coordinate system corotating with the tidally perturbed body, the longitude $\,\lambda\,$ stays constant, so we get no $\,{\bf{\dot{\rm{\lambda}}}}t\,$
 term in (\ref{116}).}
 \ba
 B_{lmpq}~=~B_{lmpq}(0)\,+~{\bf{\dot{\rm{\mbox{$B$}}}}}_{lmpq}\,t\quad,
 \label{115}
 \ea
 where
 \ba
 B_{lmpq}(0)\,\equiv\,(l\,-\,2\,p\,-\,m\,z\,')\,\omega_0\,+\,(l\,-\,2\,p\,-\,m\,z\,+\,q)\,{\cal M}_0\,-\,m\,\lambda\,-\,N\,\pi\,+\,\psi_{lm}\,~,\qquad
 \label{116}
 \ea
 \ba
 {\bf{\dot{\rm{\mbox{$B$}}}}}_{lmpq}\,\equiv~(l\,-\,2\,p\,-\,m\,z\,')\,\stackrel{\bf\centerdot}{\omega}\,+~(l\,-\,2\,p\,-\,m\,z\,+\,q)\,\stackrel{\bf\centerdot}{\cal M\,}~.\,\qquad\qquad\qquad\qquad~\qquad~\quad\qquad~\quad
 \label{117}
 \ea
 The next step is to Fourier-expand the cosines of the differences (\ref{113}), and to single out the true Fourier modes.$\,$\footnote{~The cosines of the differences (\ref{113}) enter expression (\ref{1b}) for the perturbing potential $\,W\,$. However, when the body is librating in a resonance, we
 cannot interpret these quantities' time-derivatives, $~\omega_{\textstyle{_{lmpq}}}
 \,$, as Fourier tidal modes. Indeed, the presence of a rapidly changing component in (\ref{113}) guarantees that $\,\omega_{\textstyle{_{lmpq}}}\,$, too, contains a term oscillating with the frequency $\,n\,$ comparable to $\,\omega_{\textstyle{_{lmpq}}}\,$ itself ~---~see expression (\ref{quantities}) and explanation thereafter.
 This observation makes it necessary to Fourier-expand the cosines of the differences (\ref{113}), in quest for the true Fourier modes.
 }
 Calculations in Appendix A.1 furnish us the following:
 \bs
 \ba
 \cos(\,B_{lmpq}\,-~{\cal A}~m~\sin {\cal M}\,)~=\,\sum_{s=-\infty}^{\infty}\,J_{s}(m{\cal A})~\cos(B_{lmpq}\,-\,s\,{\cal M})\,~,
 \label{118a}
 \ea
 where $\,J_{s}(x)\,$ are the Bessel functions of integer order. Recall that over not too long timescales the quantity $\,B\,$ can be linearised as in (\ref{115}). Over these timescales, expression (\ref{118a}) becomes:
 \ba
 \cos(\,B_{lmpq}\,-~{\cal A}~m~\sin {\cal M}\,)&=&\sum_{s=-\infty}^{\infty}\,J_{s}(m{\cal A})~\cos\left(B_{lmpq}(0)\,-\,s\,{\cal M}_0\,+\,(\,
  {\bf{\dot{\rm{\mbox{$B$}}}}}
 _{lmpq}\,-~s\,\stackrel{\bf\centerdot}{\cal M\,}\,)\,t\,\right)\,~,~~\quad\quad
 \label{118b}
 ~\\  \nonumber\\
 &=&\sum_{s=-\infty}^{\infty}\,J_{s}(m{\cal A})~\cos\left(B_{lmpq}(0)\,-\,s\,{\cal M}_0\,+\,\beta_{lmpqs}\,t\,\right)\,~,~~\quad
 \label{118c}
 \ea
 \label{118}
 \es
 the quantities
 \ba
 \nonumber
 \beta_{lmpqs}\,=~{\bf{\dot{\rm{\mbox{$B$}}}}}_{lmpq}\,-~s\,\stackrel{\bf\centerdot}{\cal M\,}&=&(l\,-\,2\,p\,-\,m\,z\,')\,\stackrel{\bf\centerdot}{\omega}\,+~(l\,-\,2\,p\,-\,
 m\,z\,+\,q\,-\,s)\,\stackrel{\bf\centerdot}{\cal M\,}\\
 \label{119}\\
 \nonumber
 &\approx &~(l\,-\,2\,p\,-\,
 m\,z\,+\,q\,-\,s)\,\stackrel{\bf\centerdot}{\cal M\,}
 \ea
 being the true Fourier tidal modes under forced libration.

 We see that the tidal theory for librating bodies contains not four but five integer ``quantum numbers'': $~l,\,m,\,p,\,q,\,s~$.

 The no-libration limit implies $\,{\cal{A}}\rightarrow 0\,$, in which case only the terms with $\,s=0\,$ survive in our expansions. Then we recover the regular Kaula harmonics: $\,\beta_{lmpq0}\,=\,\omega_{lmpq}\,$. Specifically, in a resonance,
 \ba
 \nonumber
 \beta_{lmpq0}\,=~{\bf{\dot{\rm{\mbox{$B$}}}}}_{lmpq}&=&(l\,-\,2\,p\,-\,m\,z\,')\,\stackrel{\bf\centerdot}{\omega}\,+~(l\,-\,2\,p\,-\,m\,z\,+\,q)\,\stackrel{\bf\centerdot}{\cal M\,}\\
 \label{120}\\
 \nonumber
 &=&(l\,-\,2\,p\,-\,m\,z\,+\,q)\,\stackrel{\bf\centerdot}{\cal M\,}\,\;.
 \ea

 For forced libration $\,\gamma\,=\,{\cal{A}}_1\,\sin{\cal{M}}\,+\,{\cal{A}}_2\,\sin{2\cal{M}}\,$ comprising two harmonics, the outcome will be similar (see Section \ref{two} below):
 \ba
 \cos(B_{lmpq} - m {\cal A}_1 \sin {\cal{M}}  - m {\cal A}_2 \sin 2 {\cal{M}}  )\;=\qquad\qquad\qquad\qquad\qquad\qquad\qquad\qquad
 \label{guest}\\
 \nonumber
 \sum_{s_1=-\infty}^{\infty}  \sum_{s_2=-\infty}^{\infty} J_{s_1 - 2 s_2} (m {\cal A}_1)\;J_{s_2} (m {\cal A}_2)\;\cos (B_{lmpq} - s_1 {\cal{M}})\;\,.\qquad\qquad
 \ea

 A more complex situation with free libration will be considered in Section \ref{free}.

 \subsection{The tide-raising potential and the induced tidal potential}

 Now that we have expanded $~\cos\left(v_{lmpq}-m(\lambda+\theta)+\psi_{lm}\right)~$ over the true Fourier modes, we should insert the result into the expression for the tide-raising potential $\,W\,$. Specifically, in equation (\ref{1b}), we should substitute $~\cos\left(v_{lmpq}-m(\lambda+\theta)+\psi_{lm}\right)~$ with the right-hand side of formula (\ref{118c}):
 \ba
 \nonumber
  W(\eRbold\,,~\erbold^{~*})~=~-~\frac{G\,M^*}{a}~\sum_{l=2}^{\infty}~\left(\,\frac{R}{a}\,\right)^{\textstyle{^l}}\sum_{m=0}^{l}~\frac{(l - m)!}{(l + m)!}~
  \left(\,2\;-\;\delta_{0m}\,\right)\;P_{{\it{l}}m}(\sin\phi)\,
  ~~~~~~~~~~~~~~~~~~~~\\
  \nonumber\\
   \sum_{p=0}^{l}F_{lmp}(\inc)\sum_{q=\,-\,\infty}^{\infty}G_{lpq}(e)
   \sum_{s=-\infty}^{\infty}\,J_{s}(m{\cal A})~\cos\left(B_{lmpq}(0)\,-\,s\,{\cal M}_0\,+\,\beta_{lmpqs}\,t\,\right)
  \,~.\qquad
 \label{121}
 \ea
 With the perturber residing in $\,{\rbold}\,^*=\,(a,\,e,\,i,\,\Omega,\,\omega,\,{\cal{M}}_0)\,$, this formula renders the perturbing potential at a surface point $\,{\Rbold}=(R,\,\phi,\,\lambda)\,$ of the tidally deformed body. The longitude $\,\lambda\,$ enters expression (\ref{116}) for $\,B_{lmpq}(0)\,$.

 A similar procedure for the secular part of the additional tidal potential of the distorted body yields:
 \ba
 \nonumber
 U(\erbold\,,\;\erbold^{\;*})&=&
 -\;\frac{G\,M^*}{a}\;\sum_{l=2}^{\infty}\;\left(\,\frac{R}{r}\,\right)^{\textstyle{^{l+1}}}
  \left(\,\frac{R}{a}\,\right)^{\textstyle{^{
  l}}}\sum_{m=0}^{l}\;\frac{({\it l} - m)!}{({\it l} + m)!}\;
  \left(\,2~-\;\delta_{0m}\,\right)\;P_{lm}(\sin\phi)~\sum_{p=0}^{\it
  l}\;F_{lmp}(\inc)~~~~\\
  \nonumber\\
  \nonumber\\
  &~&\left.~~~\right.\sum_{q=\,-\,\infty}^{\infty}\;G_{lpq}
  (e)~
  \sum_{s=-\infty}^{\infty}\,k_l~J_{s}(m{\cal A})~\cos\left(B_{lmpq}(0)\,-\,s\,{\cal M}_0\,+\,\beta_{lmpqs}\,t\,-\,\epsilon_{l}\,\right)
  ~~,~~\qquad~\,
 \label{122}
 \ea
 where the dynamical Love numbers and phase lags are functions of the true Fourier modes: $\,k_l\,=\,k_l(\beta_{lmpqs})\,$ and $\,\epsilon_l\,=\,\epsilon_l(\beta_{lmpqs})\,$.

 The formula (\ref{122}) gives the value of the incremental tidal potential in a point $\,{\rbold}=(r,\,\phi,\,\lambda)\,$ located $\,${\it{on or above}}$\,$ the surface of the deformed body ($\,r\geq R\,$). In practice, it is employed to compute the potential $\,U\,$ in a surface point ($\,r= R\,$).\vspace{3mm}

 To calculate the incremental tidal potential in a point $\,\rbold\,=\,\rbold\,^*\,$, i.e., to know the reaction of the perturber to the bulge it creates,
 we have to carry out the Kaula-like transformation again, i.e., to switch from $\,(r,\,\phi,\,\lambda)\,$ to orbital variables $\,(a,\,e,\,i,\,\Omega,\,\omega,\,{\cal{M}}_0)\,$, and to identify their values with those of the perturber. This procedure is more cumbersome than the ordinary Kaula transformation, because it also incorporates the step (\ref{118}). This calculation is explained at length in Appendix C. The resulting formulae (\ref{quid}) and (\ref{buck}) are long, and we do not repeat them here. More important are their derivatives giving the tidal torque.

 \subsection{Tidal torques caused by libration in longitude}

 Recall that $\,U(\erbold,\,\erbold^{~*})\,$ is the additional tidal potential created in an exterior point $\,\erbold\,=\,(r\,,\,\lambda\,,\,\phi)\,$ of an extended body, provided that the perturber of mass $\,M^*\,$ generating these tides is residing in an exterior point $\,\erbold^{~*}\,$. If a test particle of mass $\,M_{test}\,$ is located at $\,\erbold\,=\,(r\,,\,\lambda\,,\,\phi)\,$, the extended body acts on it with a torque $\,\vec{\bf T}\,$, while the test particle acts upon the extended body with an opposite torque $\,\vec{\cal{T}}\,=\,-\,\vec{\bf{T}}\,$. As demonstrated, e.g., in Williams \& Efroimsky (2012, Appendix A), the polar component of  $\,\vec{\cal{T}}\,$ is calculated through
 \ba
 \nonumber
 {\cal{T}}_{polar}^{\rm{^{\,(TIDE)}}}\,=~M_{test}~\frac{\partial U(\erbold,\,\erbold^{~*})}{\partial \lambda}~~.
 \label{123}
 \ea
 This is the orthogonal-to-equator component of the torque wherewith the test body located at $\,\erbold\,$ acts on the tidally perturbed extended body.

 In the special case when the test body coincides with the perturber, we have $\,M_{test}=M^*\,$ and $\,\erbold=\erbold^{\;*}\,$, and the polar component looks as
 \ba
 {\cal{T}}_{polar}^{\rm{^{\,(TIDE)}}}~=
 \left.
 ~M^*~\frac{\partial U(\erbold,\,\erbold^{\;*})}{\partial \lambda}\,
 \right|_{{{\erbold=\,\erbold^{\;*}}}}
 ~~\,.
 \label{124}
 \ea
 When the body is librating, the torque contains libration-generated components which accelerate the damping of free librations.

 Development of expression (\ref{124}) is presented in Appendix C, the resulting torque being given by equation (\ref{2222}). Here we shall present only its secular part:
 \ba
 \nonumber
 \langle \;{\cal{T}}_{polar}^{\rm{^{\,(TIDE)}}}\;\rangle
 \,=\;\frac{G\,{M^*}^{\,2}}{a}\;\sum_{l=2}^{\infty}\left(\,\frac{R}{a}\,\right)^{\textstyle{^{2l+1}}}
 \sum_{m=1}^{l}\;2\;m\;\frac{({\it l} - m)!}{({\it l} + m)!}\;\,\qquad\qquad\qquad\\
 \label{}\\
 \nonumber
 \sum_{p=0}^l\;F^{\,2}_{lmp}(i)\;\sum_{q=\,-\,\infty}^{\infty}\;G^{\,2}_{lpq}(e)\;\sum_{s=-\infty}^{\infty}J^{\,2}_{s}(m\,{\cal{A}})
     \;k_l(\beta_{lmpqs})~\sin\epsilon_l(\beta_{lmpqs})
 \,~_{\textstyle{_{\textstyle ,}}}
 \ea
 $J_{s}\,$ being the Bessel functions of the first kind.

 This torque is usually much smaller than the torque produced by the permanent triaxiality, and its influence on the dynamics is limited. However, the working of this torque is not small and can contribute to the heat budget of the librating body. We shall discuss this topic in a separate publication.

  \section{Expressions for libration with two harmonics \label{two}}

 In this section, we give the expressions for the tidal potential and torque in the case of a libration comprised by two main harmonics:
 \begin{equation}
 \gamma = {\cal A}_1 \sin {\cal{M}}  + {\cal A}_2 \sin 2 {\cal{M}}\,\;.
 \end{equation}
 We shall cite the final results, the detailed developments being provided in Appendix A.2.

 The tide-raising potential reads as
 \begin{equation}
 \begin{split}
 W(\eRbold\,,~\erbold^{~*}) & = \frac{G\,M^*}{a} \sum_{l=2}^{\infty} \left(\,\frac{R}{a}\,\right)^{\textstyle{^l}}\sum_{m=0}^{l}~\frac{(l - m)!}{(l + m)!}~
  \left(\,2  -\;\delta_{0m}\,\right)\;P_{{\it{l}}m}(\sin\phi)\;
   \sum_{p=0}^{l}  F_{lmp}(\inc) \\
   & \;\sum_{q=\,-\,\infty}^{\infty}G_{lpq}(e)~
 \sum_{s=-\infty}^{\infty}  \sum_{r=-\infty}^{\infty} J_{s - 2 r} (m {\cal A}_1) \, J_{r} (m {\cal A}_2) \cos (B_{lmpq}(0)\,-\,s\,{\cal M}_0\,+\,\beta_{lmpqs}\,t\,)
  ~~~,~\qquad\qquad\qquad
\end{split}
  \end{equation}
while the tidal potential of the distorted body is
  \begin{equation}
\begin{split}
 U(\erbold\,,\;\erbold^{\;*}) =
 -\;\frac{G\,M^*}{a}\;\sum_{l=2}^{\infty}\;\left(\,\frac{R}{r}\,\right)^{\textstyle{^{l+1}}}
  \left(\,\frac{R}{a}\,\right)^{\textstyle{^{
  l}}}\sum_{m=0}^{l}\;\frac{({\it l} - m)!}{({\it l} + m)!}\;
  \left(\,2~-\;\delta_{0m}\,\right)\;P_{lm}(\sin\phi)~\sum_{p=0}^{\it
  l}\;F_{lmp}(\inc) \\
  \sum_{q=\,-\,\infty}^{\infty}\;G_{lpq}
  (e)~k_l~
   \sum_{s=-\infty}^{\infty}  \sum_{r=-\infty}^{\infty} J_{s - 2 r} (m {\cal A}_1) \, J_{r} (m {\cal A}_2) \cos (B_{lmpq}(0)\,-\,s\,{\cal M}_0\,+\,\beta_{lmpqs}\,t\,-\,\epsilon_{l})\,\;.
 \label{eq5}
 \end{split}
  \end{equation}
 The corresponding expression for the tidal torque is:
  \begin{equation}
\begin{split}
  {\cal{T}}_{polar}~  & =~-~
 \frac{G~{M^*}^{\,2}}{a}\,~\sum_{l=2}^{\infty}\;\left(\,\frac{R}{r}\,\right)^{\textstyle{^{l+1}}}
  \left(\,\frac{R}{a}\,\right)^{\textstyle{^{
  l}}}\sum_{m=1}^{l}\;2\;m\;\frac{({\it l} - m)!}{({\it l} + m)!}\;
  P_{lm}(\sin\phi)~\sum_{p=0}^{\it  l}\;F_{lmp}(\inc) \\
 &  \sum_{q=\,-\,\infty}^{\infty}\;G_{lpq}
  (e)~k_l~
  \sum_{s=-\infty}^{\infty}  \sum_{r=-\infty}^{\infty} J_{s - 2 r} (m {\cal A}_1) \, J_{r} (m {\cal A}_2) \cos (B_{lmpq}(0)\,-\,s\,{\cal M}_0\,+\,\beta_{lmpqs}\,t\,-\,\epsilon_{l})
  ~~.~~\quad
 \end{split}
  \end{equation}
 The quantities $\,B_{lmpq}(0)\,$ and $\,\beta_{lmpqs}\,$ are identical to those emerging in the case with one main harmonic of libration.
 This formalism is easily generalised to the case of longitudinal libration containing three or more frequencies (see Appendix A.2 and references therein). Importantly, in all these cases we get only one extra ``quantum number'' $\,s\,$.

 \section{Tidal dissipation due to forced libration in longitude}

 A long calculation presented in Appendix E shows that for small-amplitude libration (i.e., for $\,m\,{\cal A}\,\ll\,1\,$) the dissipated power rate is
\begin{equation}
\begin{split}
\langle P \rangle_{\rm{tide}}  &
=
 \frac{G M^{*\,2}}{a} \sum_{l=2}^{\infty}    \bigg( \frac{R}{a} \bigg)^{2l+1}   \sum_{q^{\prime}=-\infty}^{\infty} \sum_{m=0}^l \sum_{p=0}^l \sum_{q=-\infty}^{\infty} \sum_{s=-\infty}^{\infty}
  \frac{(l-m)!}{(l+m)!} (2-\delta_{0m}) \, \\
&     \,  F_{lmp}^2 (i)  \, G_{lpq^{\prime}}(e) \, G_{lpq}(e) \, J_{(q^{\prime}-q+s)} (m {\cal{A}}) J_s (m {\cal{A}}) \,  \beta_{lmpqs} \,k_l(\beta_{lmpqs}) \sin\epsilon_l(\beta_{lmpqs})\,\;.
\label{long}
\end{split}
\end{equation}
 This formula can be approximated by substituting the Bessel functions with their asymptotic expressions for a small argument $\,m{\cal{A}}\,$, see Appendix E.

 It can be shown from the formula (\ref{long}) that in the synchronous spin-orbit state the input of the zeroth order in $\,{\cal{A}}(n)\,$ input reads as  $\;\,\frac{\textstyle 21}{\textstyle 2}\;\frac{\textstyle G\,M^{*\,2}\,R^{\,5}}{\textstyle a^6}\;n\;e^2\; k_2(n)\;\sin\epsilon_2(n)\;\,$, which coincides with the long-known expression from the standard theory by Peale \& Cassen (1978).
 With the linear and quadratic in $\,{\cal{A}}(n)\,$ inputs included, the tidally dissipated power in the 1:1 spin-orbit state reads as
 \ba
 \nonumber
 ^{(1:1)}\langle P\rangle_{\rm{tide}}&=&^{(1:1)}\langle P\rangle_{\rm{tide}}^{(main)}\,+\;^{(1:1)}\langle P\rangle_{\rm{tide}}^{(forced)}\\
 \label{}\\
 &=&\frac{\textstyle G\,M^{*\,2}\,R^{\,5}}{\textstyle a^6}\;n\,\left[\,\frac{21}{2}\,e^2\,-\,6\,e\,{\cal{A}}(n)\,+\,\frac{3}{2}\,{\cal{A}}^2(n)\,+\,\frac{3}{2}\,\sin^2 i\,\right]\,k_2(n)\;\sin\epsilon_2(n)\;\,.
 \nonumber
 \ea
 The linear in $\,{\cal{A}}(n)\,$ term is positive because in the 1:1 resonance the forced libration magnitude $\;{\cal{A}}(n)\;$ is negative, see equation (\ref{negative}).
 The libration-caused dissipation in different spin-orbit resonances will be discussed in more detail in Efroimsky (2017). There we shall show that for some moons in the solar system the libration-caused input into the tidal dissipation is leading and, in certain practical cases, greatly exceeds the ``main'' (unrelated to libration) term.

 \section{The case of free libration\label{free}}

 Capture into spin-orbit resonances takes place due to the energy loss via tidal friction. During its last-but-one steady rotation, the body loses some of its energy (and some of the angular velocity); and is left with energy barely sufficient to perform the final rotation. Over the final rotation, it looses another teeny portion of energy, and the remaining energy becomes a bit less than what is needed to complete the rotation. Then capture takes place, and the body performs its first free libration. The magnitude of this first libration is barely short of $\,\pi\,$ (with the total sweep thus being barely short of $\,2\pi\,$). Later, tidal torque damps free libration down to zero, unless nonlinearity steps in (see Footnote \ref{persistent}). The foundation for the resonant capture theory was laid by Goldreich \& Peale (1968). For modern extension and applications of this theory, see Noyelles et al. (2014) and Makarov et al. (2012).

 Consider free libration about a spin-orbit resonance $\,z\,$, at the frequency $\,\chi\,$ given by expression (\ref{publish}). Together, the main modes of the free and forced libration sum up to
 \ba
 \gamma\,=\,{\cal{A}}_1\,\sin{\cal{M}}\,+\,{\cal{A}}\,\sin(\chi t\,+\,\varphi)\;\,.
 \label{sumup}
 \ea
 Be mindful that, in order for the free libration to be sinusoidal, the condition (\ref{condi}) must hold. It does so for $\,|\,{\cal{A}}\,|\,\lesssim\,12^\circ\,$.

 The subsequent development intended to single out the Fourier modes will be slightly more involved than that in the end of Section \ref{4.1}, equation (\ref{guest}).
 \footnote{~Instead of the expression $\;\cos(B_{lmpq} - m {\cal A}_1 \sin {\cal{M}}  - m {\cal A}_2 \sin 2{\cal{M}})\;$, we now have to expand the expression
 $\;\cos\left(B_{lmpq} - m {\cal A}_1 \sin {\cal{M}}  - m {\cal D} \sin(\chi t\,+\,\varphi)\,\right)\;$. On doing this, we shall end up with not one but two new ``quantum numbers'', $\,s_1\,$ and $\,s_2\,$, see Section \ref{combined} and Appendix E.5 below. Generally, addition of each new independent libration frequency will add one more ``quantum number'', exceptional being the case where all the frequencies are integers of a certain frequency;  see Appendix A.2 and references therein.} To simplify things, suppose the magnitude of the free libration much exceeds that of the forced libration:
 \ba
 |\,{\cal A}_1\,|\,\ll\,|\,{\cal{A}}\,|\,\lesssim\,12^\circ\;\,.
 \label{}
 \ea
 This enables us to neglect the forced libration and to write: $\;\gamma = {\cal{A}}\,\sin(\chi t\,+\,\varphi)\,$. Now algebra simplifies because, instead of (\ref{guest}), we have to process the expression $\;\cos\left(B_{lmpq} - m {\cal A} \sin(\chi t\,+\,\varphi)\,\right)\;$. Then, with some minimal adjustment, our entire theory stays in force for free libration. Instead of equations (\ref{118} - \ref{119}), we now get:
 \bs
 \ba
 \cos(\,B_{lmpq}\,-~{\cal{A}}~m~\sin (\chi t\,+\,\varphi)\,)~=\,\sum_{s=-\infty}^{\infty}\,J_{s}(m{\cal{A}})~\cos(B_{lmpq}\,-\,s\,(\chi t\,+\,\varphi))\,~,
 \label{}
 \ea
 $J_{s}(x)\,$ being the Bessel functions of integer order. Over not too long timescales the quantity $\,B_{lmpq}\,$ can be linearised as was done in equation (\ref{115}); $\,$so the above expression turns into
 \ba
 \cos(\,B_{lmpq}\,-~{\cal{A}}~m~\sin (\chi t\,+\,\varphi)\,)&=&\sum_{s=-\infty}^{\infty}J_{s}(m{\cal{A}})~\cos\left(B_{lmpq}(0)\,-\,s\,\varphi\,+\,(\,
  {\bf{\dot{\rm{\mbox{$B$}}}}}_{lmpq}\,-~s\,\chi\,)\,t\,\right)\,~,~~\quad\quad
 \label{}
 ~\\  \nonumber\\
 &=&\sum_{s=-\infty}^{\infty}J_{s}(m{\cal A})~\cos\left(B_{lmpq}(0)\,-\,s\,\chi\,+\,\beta_{lmpqs}\,t\,\right)\,~,~~\quad
 \label{}
 \ea
 \label{}
 \es
 where the quantities
 \ba
 \nonumber
 \beta_{lmpqs}\,=~{\bf{\dot{\rm{\mbox{$B$}}}}}_{lmpq}\,-~s\,\chi&=&(l\,-\,2\,p\,-\,m\,z\,')\,\stackrel{\bf\centerdot}{\omega}\,+~(l\,-\,2\,p\,-\,
 m\,z\,+\,q)\,\stackrel{\bf\centerdot}{\cal M\,}\,-\,s\,\chi\\
 \label{trit}\\
 \nonumber
 &\approx &~(l\,-\,2\,p\,-\,
 m\,z\,+\,q)\,\stackrel{\bf\centerdot}{\cal M\,}\,-\,s\,\chi
 \ea
 are playing the role of the Fourier tidal modes.

 As demonstrated in Appendix E.4, in this situation the time-averaged dissipation rate is furnished by an expression similar to equation (\ref{long}), though with $\,q^{\,\prime}=q\;$:
\begin{equation}
\begin{split}
\langle P \rangle &
=
 \frac{G M^{*\,2}}{a} \sum_{l=2}^{\infty}    \bigg( \frac{R}{a} \bigg)^{2l+1}    \sum_{m=0}^l \sum_{p=0}^l \sum_{q=-\infty}^{\infty} \sum_{s=-\infty}^{\infty}
  \frac{(l-m)!}{(l+m)!} (2-\delta_{0m}) \, \\~\\
&     \,  F_{lmp}^{\,2} (i)  \, \, G_{lpq}^2(e) \, J_s^2 (m {\cal{A}}) \,  \beta_{lmpqs} \,k_l(\beta_{lmpqs}) \sin\epsilon_l(\beta_{lmpqs})\,\;.
\label{dit}
\end{split}
\end{equation}

 Be mindful that
 our treatment of the problem is exact, in that
 we did not develop the tidal deformation or potential in the leading order over the libration magnitude, as was done in Correia et al. (2014, Section 6). At the same time, we relied on the condition (\ref{condi}) of smallness of $\,\gamma(t)\,$, in order to expand it into a Fourier time series (\ref{eq1}); hence the limitation on the libration magnitude within our approach, $\,|\,
 \gamma
 \,|\,\lesssim\,12^\circ\,$.

 \section{The case of free libration superimposed on forced libration\label{combined}}

 If we now wish to keep both terms in the expression (\ref{sumup}) then, instead of equations (\ref{118} - \ref{119}), we obtain the following:
 \ba
 \nonumber
 \cos(\,B_{lmpq}\,-~{\cal{A}}_1~m~\sin{\cal{M}}~-~{\cal{A}}~m~\sin (\chi t\,+\,\varphi)\,)  \qquad\qquad\qquad\qquad   ~\\
 \label{}\\
 \nonumber
 =\,\sum_{s_1=-\infty}^{\infty}\,\sum_{s_2=-\infty}^{\infty}\,J_{s_1}(m{\cal{A}}_1)\,J_{s_2}(m{\cal{A}})~\cos(B_{lmpq}\,-\,s_1\,{\cal{M}}\,-\,s_2\,(\chi t\,+\,\varphi))\,\;,
 \ea
 which can be easily proven with aid of formula (\ref{expansion}) from Appendix A.1.

 As explained in Section \ref{interpretation}, the argument of the cosine can be linearised over timescales shorter than the period of the apsidal precession:
 \ba
 \nonumber
 && \cos(\,B_{lmpq}\,-~{\cal{A}}_1~m~\sin{\cal{M}}~-~{\cal{A}}~m~\sin (\chi t\,+\,\varphi)\,)  \qquad\qquad\qquad\qquad   ~\\
 \nonumber\\
 &=&\sum_{s_1=-\infty}^{\infty}\,\sum_{s_2=-\infty}^{\infty}\,J_{s_1}(m{\cal{A}}_1)\,J_{s_2}(m{\cal{A}})~\cos\left(B_{lmpq}(0)\,-\,s_1\,{\cal{M}}_0\,-\,s_2\,\varphi\,+\,
 (\dot{B}_{lmpq}\,-\,s_1\,n\,-\,s_2\,\chi)\,t\,\right)
 \nonumber\\
 &=&\sum_{s_1=-\infty}^{\infty}\,\sum_{s_2=-\infty}^{\infty}\,J_{s_1}(m{\cal{A}}_1)\,J_{s_2}(m{\cal{A}})~\cos\left(B_{lmpq}(0)\,-\,s_1\,{\cal{M}}_0\,-\,s_2\,\varphi\,+\,
 \beta_{lmpqs_1s_2}\,t\,\right)\,\;,
 \label{labour}
 \ea
 the quantities
 \ba
 \nonumber
 \beta_{lmpqs_1s_2}&=&{\bf{\dot{\rm{\mbox{$B$}}}}}_{lmpq}\,-~s_1\,n\,-\,s_2\,\chi
 ~\\
 \nonumber
 &=&(l\,-\,2\,p\,-\,m\,z\,')\,\stackrel{\bf\centerdot}{\omega}\,+~(l\,-\,2\,p\,-\,
 m\,z\,+\,q\,-\,s_1)\,n\,-\,s_2\,\chi\\
 &\approx &~(l\,-\,2\,p\,-\,
 m\,z\,+\,q\,-\,s_1)\,n\,-\,s_2\,\chi
  \label{six}
 \ea
 being the Fourier tidal modes under forced libration combined with free libration. These modes have six indices.

 Employment of these formulae to calculate the power dissipated by the superposition of forced and free libration entails calculations more cumbersome than in the case of only forced or only free libration. Besides a longer expression for a tidal mode, we shall get in the answer cross terms containing cross products like
 $\;J_{q^{\,\prime}-q+s_1}(m {\cal{A}}_1)\,J_{s_1}(m {\cal{A}}_1)\,J^2_{s_2}(m {\cal{A}})\;$. It is however explained in Appendix E.5 that, in case we agree to consider the terms at most quadratic in the libration magnitudes, then the cross terms may be neglected, and we shall keep only the terms containing $\;J_{q^{\,\prime}-q+s_1}(m {\cal{A}}_1)\,J_{s_1}(m {\cal{A}}_1)\;$ and those containing $\;J^2_{s_2}(m {\cal{A}})\;$. As a result of this, the total power will be a sum of the forced-libration power (\ref{long}), with the five-indexed tidal mode $\,\beta_{lmpqs}\,$ given by expression (\ref{119}), and the free-libration power (\ref{dit}), with the five-indexed tidal mode $\,\beta_{lmpqs}\,$ given by expression (\ref{trit}).

 \section{Conclusions}

 In the article thus far, we have demonstrated that the spectrum of Fourier modes of libration-caused tides differs from the spectrum of tidal modes ensuing from the Darwin-Kaula theory. It has turned out that the conventional formulae from the Darwin-Kaula theory are applicable only to nonresonant spin (or to resonant spin with no physical libration). This motivated us to generalise the Darwin-Kaula theory and to derive the spectrum of the tidal Fourier modes in a librating rotator. We carried out this calculation for the libration angle comprising one or two forced harmonics, and we also explained how to generalise this calculation to the case of free libration.

 The presence of one libration mode adds one more index, $\,s\,$, in the numbering of the tidal modes. The new set of modes now becomes $\,\beta_{lmpqs}\,$ and incorporates the conventional modes $\,\omega_{lmpq}\,$ obtained from the theory of Kaula (1964) with no libration: $\,\beta_{lmpq0}\,=\,\omega_{lmpq}\,$. Interestingly, only one additional ``quantum number'' emerges in the theory, no matter how many forced harmonics $\,\sin j{\cal{M}}\,$ are contained in the libration spectrum (this observation, generally, not being true for free libration harmonics).

 We have derived an expression for the polar tidal torque acting on a librating body. This torque accelerates damping of free libration, and generates a small correction to libration.

 We also have calculated the tidal input  \footnote{~Here we say $\,${\it{tidal\, input}}, because under libration the energy is damped also by the alternating parts of the centripetal and toroidal forces emerging in the rotator. For a detailed investigation on the energy dissipation in a librating body, see our forthcoming work Efroimsky (2017).} into the dissipation rate in a rotator librating in an arbitrary spin-orbit resonance. In the absence of libration, our expression coincides with that for the power exerted by tides in a body steadily spinning in an appropriate resonance.

 Finally, we considered a situation where the principal forced libration mode is superimposed with a free libration mode. We have demonstrated that, to a very good approximation, the tidally dissipated power can in this case be presented as a sum of two independent inputs: the power dissipated by forced libration (calculated in neglect of the free libration) and the power dissipated by the free libration (calculated in neglect of the forced libration).

 \section*{Acknowledgments}

 The authors are grateful to Val\'ery Lainey and Valeri V. Makarov for extremely helpful and stimulating discussions on the topic of this work.
 The authors' special thanks go to the Reviewers, Gwenaël Boué
 and Beno\^it Noyelles, whose incisive questions and comments were very instrumental in improving the quality of the paper.
 ME would also like to thank Konstantin V. Kholshevnikov for his highly valuable consultations on the theory of gravitational potential.
 This research has made use of NASA's Astrophysics Data System.


 \section*{\underline{\textbf{\Large{Appendix $\,$A.}}}
 \vspace{3mm}~\\
 ~~~\Large{Fourier expansion of the perturbing potential.\vspace{1mm}\\
 The case of resonant spin states}}

 \subsection*{A.1~~~Libration with one harmonic\label{shpaga}}

 Our goal here is to Fourier-expand the expression $~\cos(\,B_{lmpq}\,-~{\cal A}~m~\sin {\cal M}\,)~$ showing up in an $\,lmpq\,$ term of the Darwin-Kaula expansion (\ref{1b})
 for the perturbing potential $\,W\,$. We begin with the expansion (Abramovitz \& Stegun 1972, eqns. 9.1.42 - 9.1.43)
 \ba
 \exp(i\,x\,\sin y)\;=\;\sum_{s\,=\,-\,\infty}^{\infty}\,J_s(x)\;\exp(isy)\,\;,
 \label{expansion}
 \ea
 where it is implied that $\,x=m{\cal{A}}\,$ and $\,y=\,-\,{\cal{M}}\,$, and where the Bessel functions of a negative order are defined through $\,~J_{\textstyle{_{\,-\,s}}}\,\equiv~
 (-1)^{\textstyle{^s}}\,J_{\textstyle{_{s}}}\,~,~~s\in{\cal N}\,~$.

 Multiplying the expansion (\ref{expansion}) by $\,\exp iB_{lmpq}\,$ we arrive at the relation
 \ba
 \exp i(B_{lmpq}\,-\,m\,{\cal{A}}\,\sin {\cal{M}})\;=\;\sum_{s\,=\,-\,\infty}^{\infty}\,J_s(x)\;\exp i(B_{lmpq}\,-\,s\,{\cal{M}})\,\;,
 \label{}
 \ea
 its real part being
    \ba
    \cos(\,B_{lmpq}\,-m~{\cal A}~\sin {\cal M}\,)~=\,\sum_{s=-\infty}^{\infty}\,J_{s}(m{\cal A})~\cos(B_{lmpq}\,-\,s\,{\cal M})\,~.
    \label{A6c}
    \ea
 Noting that $\,J_0 (0) = 1\,$ and that $\,J_n (0) = 0\,$ for $\,n\neq 0\,$, we recover the original expression for $\,\cos B_{lmpq}\,$ in the no-libration case with $\,{\cal A} = 0\,$.

\subsection*{A.2~~~Libration with multiple harmonics\label{Dattoli}}

 Generalisation of the above development to the case of libration containing multiple harmonics,
\begin{equation}
\cos(B_{lmpq} - m {\cal A}_1 \sin {\cal{M}} - m {\cal A}_2 \sin 2 {\cal{M}}   - ...)\;\,,
\end{equation}
 is made easier by considering a 2-dim version of the generalised Jacobi-Anger expansion
\begin{equation}
e^{i(x \sin \gamma + y \sin 2 \gamma)}  = \sum_{n=-\infty}^{\infty}  \, ^{(2)}J_n (x,y)\;e^{i n \gamma}\,\;.
\label{eq3}
\end{equation}
 This expansion involves the generalised Bessel functions
\begin{equation}
 ^{(2)}J_n (x,y)\;= \sum_{k=-\infty}^{\infty} J_{n-2k} (x)\; J_k (y)\,\;,
 \label{eq4}
\end{equation}
 see Dattoli et al. (1996, 1998), Korsch et al. (2006), and references therein. After some trigonometric manipulations, we arrive at a generalisation of equation (\ref{A6c}):
 \ba
 \cos(B_{lmpq} - m {\cal A}_1 \sin {\cal{M}}  - m {\cal A}_2 \sin 2 {\cal{M}}  )\;=
 \qquad\qquad\qquad\qquad\qquad\qquad\qquad
 \nonumber\\
 \label{}\\
 \nonumber
 \sum_{s_1=-\infty}^{\infty}  \sum_{s_2=-\infty}^{\infty} J_{s_1 - 2 s_2} (m {\cal A}_1) J_{s_2} (m {\cal A}_2) \cos (B_{lmpq} - s_1 {\cal{M}})\,\;.
 \label{generalisation}
 \ea
 This coincides with equation (\ref{A6c}) in the limit of $\,{\cal A}_2 = 0\,$.

 Generalisation to the case of libration with $\,N\,$ harmonics of the same frequency is straightforward owing to the availability of a generalised version of our formulae (\ref{eq3} - \ref{eq4}):
 \begin{equation}
 \exp \bigg( i \sum_{s=1}^{N} x_s \sin s\gamma \bigg)\;= \sum_{n = -\infty}^{\infty} \, ^{(N\,,\,...\,,2)}J_n(x_1\,,\,...\,,\,x_N)\;e^{i n \gamma}
 \end{equation}
 where the generalised Bessel functions are defined recursively via
 \begin{equation}
 ^{(N\,,\,N-1\,,\,...\,,\,2)}J_n(x_1\,,\,...\,,\,x_N)\;=\sum_{l=-\infty}^{\infty} \, ^{(N-1,...,2)} J_{n-Nl}(x_1\,,\,...\,,\,x_{N-1})\;J_l (x_N)\,\;,
 \end{equation}
 see Dattoli et al. (1996, eqns 8 - 9).

 \section*{\underline{\textbf{\Large{Appendix $\,$B.}}}\vspace{3mm}\\
 ~~~\Large{The tidal theory of Kaula (1964), with no libration}}\label{AppB}

 We have to recall in short some developments from the Kaula theory of tides, as a preparation for generalisation thereof to the case with libration. Referring the reader for a more extensive discussion to Efroimsky \& Makarov (2013) and Efroimsky (2015), here we provide only several key formulae. As ever, we equip with asterisk all variables related to the perturber, while those with no asterisk pertain to a location where the potentials are measured.

 \subsection*{B.1~~~Kaula's trigonometric formula\label{B1}}

 Kaula (1961) derived a trigonometric formula interconnecting the spherical coordinates $\,(r^*,\,\phi^*,\,\lambda^*)\,$ of an orbiter with its Keplerian elements $\,(a^*,\,e^*,\,i^*,\,\Omega^*,\,\omega^*,\,{\cal{M}}_0^*)\,$ introduced in a frame which is associated with the equator but is not co-rotating with it:
 \ba
 \nonumber
 \left(\,\frac{1}{r^{\,*}}\,\right)^{{\it l}+1}P_{lm}(\sin
 \phi^*)\;\left[\;\cos m\lambda^*\;+\;\sqrt{-1}\;\sin m\lambda^*\;\right]
 ~=~~~~~~~~~~~~~~~~~~~~~~~~~~~~~~~~~~~~~~~~~~~~~~~~~~~~~~~~~~~~~~~\\
 \label{kaula}
 \ea
 \ba
 \nonumber
 \left(\frac{1}{a^{\,*}}\right)^{{\it
 l}+1}\sum_{p=0}^{\infty}F_{{\it l}mp}(\inc^*)\sum_{q=\,-\,\infty}^{\infty}
 G_{{\it l}pq}(e^*)\;
  \left\{
  \begin{array}{c}
  \cos \left(\,
  v_{{\it l}mpq}^*\,-\,m\,\theta^*\,\right)\,+\,\sqrt{-1}\;\sin \left(\,
  v_{{\it l}mpq}^*\,-\,m\,\theta^*\,\right)  \\
  \sin \left(\,
  v_{{\it l}mpq}^*\,-\,m\,\theta^*\,\right)\,-\,\sqrt{-1}\;\cos \left(\,
  v_{{\it l}mpq}^*\,-\,m\,\theta^*\,\right)
  \end{array}
  \right\}^{{\it l}\,-\,m\;\;
  \mbox{\small even}}_{{\it l}\,-\,m\;\;\mbox{\small odd}~~~~~~~{\textstyle
  ,}}~~~~
  \ea
 where the auxiliary combinations $\,v_{{\it l}mpq}^*\,$ are defined as
 \ba
 v_{lmpq}^*\;\equiv\;(l-2p)\omega^*\,+\,(l-2p+q){\cal M}^*\,+\,m\,\Omega^*~~~,
  \ea
 while the notation $\,\sqrt{-1}\,$ is used to avoid confusion with the inclination.

 \subsection*{B.2~~~From the spherical coordinates to the orbital variables. Step 1\label{B2}}

 In a surface point $\,\Rbold\,=\,(R,\,\phi,\,\lambda)\,$ of a tidally perturbed body, the potential due to a perturber located in
 $\,\rbold^{\,*}=\,
 (r^{\,*},\,\phi^*,\,\lambda^*)
 \,$ can be written as
 \bs
 \ba
 W(\eRbold,\,\erbold^{\;*})\,=\;-\;G\;M^*\,\left[\frac{1}{|\eRbold
 -\erbold^{\;*}|}\;-\;\frac{{\eRbold}\,\cdot\,{\erbold}^{\;*}}{|
 \erbold^{\;*}|^3}   \right]\;
 =\;-\;\frac{G\;M^*}{r^{\,*}}\,
 \sum_{{\it{l}}=2}^{\infty}\,\left(\,\frac{R}{r^{\;*}}\,\right)^{
 \textstyle{^{\it{l}}}}\,P_{\it{l}}(\cos \gamma)\;\;
 \ea
  \ba
 =\,-\,\frac{G\;M^*}{r^{\,*}}
 \sum_{{\it{l}}=2}^{\infty}\left(\frac{R}{r^{\;*}}\right)^{
 \textstyle{^{\it{l}}}}\sum_{m=0}^{\it l}\frac{({\it l} - m)!
 }{({\it l} + m)!}(2-\delta_{0m})P_{{\it{l}}m}(\sin\phi)P_{{
 \it{l}}m}(\sin\phi^*)\;\cos m(\lambda-\lambda^*)\,\;.
 \ea
 \es
 Applying his trigonometric formula to the variables with asterisk (those related to the perturber), Kaula (1964) wrote down the perturbing potential as
 \bs
 \ba
 \nonumber
  W(\eRbold\,,\;\erbold^{\;*})\;=\;-\;
  \frac{G\,M^*}{a^*}\;\sum_{{\it
  l}=2}^{\infty}\;\left(\,\frac{R}{a^*}\,\right)^{\textstyle{^{\it
  l}}}\sum_{m=0}^{\it l}\;\frac{({\it l} - m)!}{({\it l} + m)!}\;
  \left(\,2  \right. ~~~~~~~~~~~~~~~~~~~~~~~~~~~~~~~~~~~~~~~~~~~~~~~~~~~~~~~\\
                                   \nonumber\\
                                   \nonumber\\
       \left.
  ~~~ -\;\delta_{0m}\,\right)\;P_{{\it{l}}m}(\sin\phi)\;\sum_{p=0}^{\it
  l}\;F_{lmp}(\inc^*)\;\sum_{q=\,-\,\infty}^{\infty}\;G_{{\it l}pq}
  (e^*)
  \left\{
  \begin{array}{c}
   \cos   \\
   \sin
  \end{array}
  \right\}^{{\it l}\,-\,m\;\;
  \mbox{\small even}}_{{\it l}\,-\,m\;\;\mbox{\small odd}} \;\left(
  v_{{\it l}mpq}^*-m(\lambda+\theta^*)  \right)
 ~~~.~~~~~~~~~
 \label{cogito}
 \ea
 More convenient is the equivalent expression
 \ba
 \nonumber
  W(\eRbold\,,~\erbold^{~*})\;=~-~\frac{G\,M^*}{a^*}~\sum_{l=2}^{\infty}~\left(\,\frac{R}{a^*}\,\right)^{\textstyle{^l}}\sum_{m=0}^{l}~\frac{(l - m)!}{(l + m)!}~
  \left(\,2
  \right. ~~~~~~~~~~~~~~~~~~~~~~~~~~~~~~~~~~~~~~\\  \nonumber\\  \left.~~~
   -\;\delta_{0m}\,\right)\;P_{lm}(\sin\phi)\;\sum_{p=0}^{l}F_{lmp}(\inc^*)\;\sum_{q=\,-\,\infty}^{\infty}G_{lpq}(e^*)
 ~\cos\left(v_{lmpq}^*-m(\lambda+\theta^*)+\psi_{lm}\right)~~~,~\qquad\qquad\qquad
 \label{ergo}
 \ea
 \label{sum}
 \es
 where supplementary phases are given by
 \ba
 \psi_{lm}~=~\left[\,(\,-\,1)^{\,l-m}\,-\,1\,\right]~\frac{\pi}{4}\;~.~\qquad~ ~\qquad~ ~\qquad~
 \ea

 Within a static tidal problem, a corresponding expression for the additional tidal potential of the disturbed body, observed in an exterior point $\,\rbold\,=\,(r,\,\phi,\,\lambda)\,$ with $\,r\geq R\,$, is:
 \ba
 \nonumber
 U(\erbold\,,\;\erbold^{\;*})&=&\sum_{{\it l}=2}^{\infty}~U_{\it{l}}(\erbold)~=~\sum_{l=2}^{\infty}~k_{\it
 l}\;\left(\,\frac{R}{r}\,\right)^{l+1}\;W_{\it{l}}(\eRbold\,,\;\erbold^{\;*})
    ~\\ \nonumber\\  \nonumber
 &=&
 -\;\frac{G\,M^*}{a}\;\sum_{l=2}^{\infty}\;\left(\,\frac{R}{r}\,\right)^{\textstyle{^{l+1}}}
  \left(\,\frac{R}{a}\,\right)^{\textstyle{^{
  l}}}\sum_{m=0}^{l}\;\frac{({\it l} - m)!}{(l + m)!}\;
  \left(\,2~-\;\delta_{0m}\,\right)\;P_{lm}(\sin\phi)~~~~\\
  \nonumber\\
  &~&\left.~\right.\sum_{p=0}^{\it
  l}\;F_{lmp}(\inc^*)\;\sum_{q=\,-\,\infty}^{\infty}\;G_{lpq}
  (e^*)~k_l~
  \cos\left(\,v^*_{lmpq}-\,m\,(\lambda~+~\theta^*)~+~\psi_{lm}~-~\epsilon_l
  \, \right)\;~,~~\qquad
 \label{}
 \ea
 $k_l\,$ being the $\,${\it{static}}$\,$ Love numbers.

 In the case of evolving tides, however, each Fourier term acquires a phase lag of its own:
  \ba
 \nonumber
 U(\erbold\,,\;\erbold^{\;*})
 &=&
 -\;\frac{G\,M^*}{a}\;\sum_{l=2}^{\infty}\;\left(\,\frac{R}{r}\,\right)^{\textstyle{^{l+1}}}
  \left(\,\frac{R}{a}\,\right)^{\textstyle{^{
  l}}}\sum_{m=0}^{l}\;\frac{({\it l} - m)!}{(l + m)!}\;
  \left(\,2~-\;\delta_{0m}\,\right)\;P_{lm}(\sin\phi)~~~~\\
  \nonumber\\
  &~&\left.~\right.\sum_{p=0}^{\it
  l}\;F_{lmp}(\inc^*)\;\sum_{q=\,-\,\infty}^{\infty}\;G_{lpq}
  (e^*)~k_l~
  \cos\left(\,v^*_{lmpq}-\,m\,(\lambda~+~\theta^*)~+~\psi_{lm}~-~\epsilon_l
  \, \right)\;~,~~\qquad
 \label{bach}
 \ea
 $\,k_l\,$ and $\,\epsilon_l\,$ being the $\,${\it{dynamical}}$\,$ Love numbers and the phase lags. As functions of the tidal Fourier modes, they are numbered with the degree $\,l\,$ only. However, the values of $\,k_l\,$ and $\,\epsilon_l\,$ depend also on $\,m\,$, $\,p\,$, $\,q\,$ because the Fourier modes depend upon all four indices. We shall discuss this in Appendix B.4 below.

 Now suppose that the test particle coincides with the perturber, so $\,\rbold=\rbold^{\;*}\,$. Then the negative of the derivative of the above expression with respect to the longitude $\,\lambda\,$, multiplied by the test particle mass, renders the polar component of the orbital torque acting on the test particle. The negative of that torque (i.e., simply the derivative of $\,U\,$ over $\,\lambda\,$, multiplied by the test particle mass) will be the polar torque acting on the tidally perturbed body:
 \ba
 \nonumber
 {\cal{T}}_{polar}^{\rm{^{\,(TIDE)}}}&=&
 -\;\frac{G\,{M^*}^{\,2}}{a}\;\sum_{l=2}^{\infty}\;\left(\,\frac{R}{r}\,\right)^{\textstyle{^{l+1}}}
  \left(\,\frac{R}{a}\,\right)^{\textstyle{^{
  l}}}\sum_{m=0}^{l}\;2\;m\;\frac{(l - m)!}{(l + m)!}\;
  \;P_{lm}(\sin\phi)~~~~\\
  \nonumber\\
  &~&\left.~\right.\sum_{p=0}^{\it
  l}\;F_{lmp}(\inc)\;\sum_{q=\,-\,\infty}^{\infty}\;G_{lpq}
  (e)~k_l~
  \sin\left(\,v^*_{lmpq}-\,m\,(\lambda~+~\theta^*)~+~\psi_{lm}~-~\epsilon_l
  \, \right)\;~.~~\qquad
 \label{mozart}
 \ea

 \subsection*{B.3~~~From the spherical coordinates to the orbital variables. Step 2\label{B3}}

 To process expression (\ref{bach}) for the potential, we rewrite the cosine from that equation as
 \ba
 \nonumber
 \cos\left(\,v^*_{lmpq}-\,m\,(\lambda+\theta^*)~+~\psi_{lm}~-~\epsilon_l\, \right)~=\qquad\qquad\qquad\qquad\qquad\qquad\qquad\qquad\qquad\qquad\;\;\;\;\quad\\
   \label{kilo}\\
 \nonumber
 \cos\left(\,v^*_{lmpq}-\,m\,\theta^*~+~\psi_{lm}~-~\epsilon_l\, \right)~\cos m\lambda\;+
 \sin\left(\,v^*_{lmpq}-\,m\,\theta^*~+~\psi_{lm}~-~\epsilon_l\, \right)~\sin m\lambda\qquad~
 \ea
 and apply Kaula's trigonometric formula to the variables with no asterisk (those related to the position of the test particle):
 \ba
 \left(\frac{R}{r}\right)^{l+1}\,P_{lm}(\sin\phi)\;\cos m\lambda\;=\,\left(\frac{R}{a}\right)^{l+1}\,\sum_{h=0}^l F_{lmh}(\inc)\sum_{j=-\infty}^{\infty}
  G_{lhj}(e)\;\cos(v_{lmpq}-\,m\,\theta~+~\psi_{lm})\;,\;\;\;
 \label{giga}
 \ea
  \ba
 \left(\frac{R}{r}\right)^{l+1}\,P_{lm}(\sin\phi)\;\sin m\lambda\;=\,\left(\frac{R}{a}\right)^{l+1}\,\sum_{h=0}^l F_{lmh}(\inc)\sum_{j=-\infty}^{\infty}
  G_{lhj}(e)\;\sin(v_{lmpq}-\,m\,\theta~+~\psi_{lm})\;.\;\;\;
 \label{mega}
 \ea
  Combined, the above three equations entail:
 \ba
 \nonumber
 \left(\frac{R}{r}\right)^{l+1}\,P_{lm}(\sin\phi)\;\cos\left(\,v^*_{lmpq}-\,m\,(\lambda+\theta^*)~+~\psi_{lm}~-~\epsilon_l\, \right)~=\\
 \nonumber\\
  \left(\frac{R}{a}\right)^{l+1}\,\sum_{h=0}^{\it l}F_{lmh}(\inc)\sum_{j=-\infty}^{\infty}
  G_{lhj}(e)\;\cos\left(\,
  \left[v_{{\it l}mpq}^*-\,m\theta^*\,\right]
  - \left[v_{{\it l}mhj}-m\theta\,\right]\,-\,\epsilon_{l}\,\right)
 \,~_{\textstyle{_{\textstyle .}}}
 \label{}
 \ea
 Insertion thereof into formula (\ref{bach}) results in
 \ba
 \nonumber
 U(\erbold,\,\rbold^{\,*})\;=\;-\;\sum_{l=2}^{\infty}\;\left(\,\frac{R}{a}\,\right)^{\textstyle{^{{\it
  l}+1}}}\frac{G\,M^*}{a^*}\;\left(\,\frac{R}{a^*}\,\right)^{\textstyle{^l}}\sum_{m=0}^{\it l}\;
  \frac{({\it l} - m)!}{({\it l} + m)!}\;\left(\,2\;\right.~~~~~~~~~~~~~~~~~~~~~\\
                                                                                                         \nonumber\\
                                    \nonumber
 ~~~~\left.-\,\delta_{0m}\right)\sum_{p=0}^{l}F_{lmp}(\inc^*)\sum_{q=-\infty}^{\infty}G_{lpq}
  (e^*) \sum_{h=0}^{\it l}F_{lmh}(\inc)\sum_{j=-\infty}^{\infty}
  G_{lhj}(e)~~~~~~~\\
  \nonumber\\
  k_{\it l}\;\cos\left(\,
  \left[v_{{\it l}mpq}^*-\,m\theta^*\,\right]
  - \left[v_{{\it l}mhj}-m\theta\,\right]\,-\,\epsilon_{l}\,\right)
 \,~_{\textstyle{_{\textstyle .}}}~~~~
 \label{}
 \ea
As was pointed out in Efroimsky (2012a), in this expansion the terms with $\,h=p,\,j=q\,$ form the secular part of the potential, the other terms being oscillatory.

 When the test particle coincides with the perturber, the secular part of this expression becomes simply
  \ba
 \nonumber
 U(\erbold^{\,*},\,\rbold^{\,*})\;=\;-\;\sum_{l=2}^{\infty}\;\left(\,\frac{R}{a}\,\right)^{\textstyle{^{{\it
  l}+1}}}\frac{G\,M^*}{a^*}\;\left(\,\frac{R}{a^*}\,\right)^{\textstyle{^l}}\sum_{m=0}^{\it l}\;
  \frac{({\it l} - m)!}{({\it l} + m)!}\;\left(\,2\;\right.~~~~~~~~~~~~~~~~~~~~~\\
                                                                                                         \label{}\\
                                    \nonumber
 ~~~~\left.-\,\delta_{0m}\right)\sum_{p=0}^{l}F_{lmp}(\inc^*)\sum_{q=-\infty}^{\infty}G_{lpq}
  (e^*) \sum_{h=0}^{\it l}F_{lmh}(\inc)\sum_{j=-\infty}^{\infty}
  G_{lhj}(e)\;
  k_{\it l}\,\cos\epsilon_{l}
 \,~_{\textstyle{_{\textstyle .}}}~~~~
 \ea
 The expression (\ref{mozart}) for the tidal torque is processed similarly. When the test particle coincides with the perturber, the secular part of the polar torque is
  \ba
 \nonumber
 {\cal{T}}_{polar}^{\rm{^{\,(TIDE)}}}\;=\;-\;\sum_{l=2}^{\infty}\;\left(\,\frac{R}{a}\,\right)^{\textstyle{^{{\it
  l}+1}}}\frac{G\,{M^*}^{\,2}}{a^*}\;\left(\,\frac{R}{a^*}\,\right)^{\textstyle{^l}}\sum_{m=1}^{\it l}\;2\;m\;
  \frac{({\it l} - m)!}{({\it l} + m)!}~~~~~~~~~~~~~~~~~~~~~\\
                                                                                                         \label{torque}\\
                                    \nonumber
 ~~~~\sum_{p=0}^{l}F_{lmp}(\inc^*)\sum_{q=-\infty}^{\infty}G_{lpq}
  (e^*) \sum_{h=0}^{\it l}F_{lmh}(\inc)\sum_{j=-\infty}^{\infty}
  G_{lhj}(e)\;
  k_{\it l}\,\sin\epsilon_{l}
 \,~_{\textstyle{_{\textstyle .}}}~~~~
 \ea
 In the above expansions, it is implied that  $\,k_l\,=\,k_l(\omega_{lmpq})\,$ and $\,\epsilon_l\,=\,\epsilon_l(\omega_{lmpq})\,$, with
 \ba
 \omega_{lmpq}\,=\,(l-2p)\,\dot{\omega}\,+\,(l-2p+q)\dot{\cal M}\,+\,m\,(\dot{\Omega}\,-\,\dot{\theta})
 \label{}
 \ea
 being the tidal Fourier modes, as we saw in Section \ref{interpretation}.

 \section*{\underline{\textbf{\Large{Appendix $\,$C.}}}\vspace{3mm}\\
 ~~~\Large{Generalisation of Kaula's theory to the case with libration}\label{C}}

 When a spinning body is caught in a spin-orbit resonance, its rotation angle $\,{\theta}\,$ can be split into longitudinal libration $\,\gamma\,$ and a constant part $\,\theta_{res}\,$:
 \ba
 \theta\;=\;\theta_{res}\;+\;\gamma\;=\;\theta_{res}\;+\;m\;{\cal{A}}\;\sin{\cal{M}}\,\;.
 \label{ged}
 \ea
 For details, see Section \ref{small} and, specifically, equation (\ref{defff}). At this point, we do not need a specific expression for $\,\theta_{res}\,$, while the libration $\,\gamma\,$ is approximated with its leading harmonic.

 In the presence of libration, most derivations from Appendix B remain in force, except that equations (\ref{giga}) and (\ref{mega}) must be now
 processed with aid of formulae (\ref{118a}) and (\ref{ged}):
  \ba
  \nonumber
 \left(\frac{R}{r}\right)^{l+1}\,P_{lm}(\sin\phi)\;\cos m\lambda\;=\,\left(\frac{R}{a}\right)^{l+1}\,\sum_{h=0}^l F_{lmh}(\inc)\sum_{j=-\infty}^{\infty}
  G_{lhj}(e)\;\cos(v_{lmpq}-\,m\,\theta~+~\psi_{lm})\;=\\
  \nonumber\\
  \nonumber
  \left(\frac{R}{a}\right)^{l+1}\,\sum_{h=0}^l F_{lmh}(\inc)\sum_{j=-\infty}^{\infty}
  G_{lhj}(e)\;\cos(v_{lmpq}-\,m\,\theta_{res}~+~\psi_{lm}\;-\;m\;{\cal{A}}\;\sin{\cal{M}})\;=\\
  \left(\frac{R}{a}\right)^{l+1}\,\sum_{h=0}^l F_{lmh}(\inc)\sum_{j=-\infty}^{\infty}G_{lhj}(e)\;
  \sum_{s_1=\,-\,\infty}^{\infty}J_{s_1}(m\,{\cal{A}})\,\cos(v_{lmpq}-\,m\,\theta_{res}~+~\psi_{lm}~-~s_1\,{\cal{M}})
  \;,\;\;\;
 \label{giga2}
 \ea
  \ba
  \nonumber
 \left(\frac{R}{r}\right)^{l+1}\,P_{lm}(\sin\phi)\;\sin m\lambda\;=\,\left(\frac{R}{a}\right)^{l+1}\,\sum_{h=0}^l F_{lmh}(\inc)\sum_{j=-\infty}^{\infty}
  G_{lhj}(e)\;\sin(v_{lmpq}-\,m\,\theta~+~\psi_{lm})\;=\\
  \nonumber\\
  \nonumber
  \left(\frac{R}{a}\right)^{l+1}\,\sum_{h=0}^l F_{lmh}(\inc)\sum_{j=-\infty}^{\infty}
  G_{lhj}(e)\;\sin(v_{lmpq}-\,m\,\theta_{res}~+~\psi_{lm}\;-\;m\;{\cal{A}}\;\sin{\cal{M}})\;=\\
  \nonumber\\
  \left(\frac{R}{a}\right)^{l+1}\,\sum_{h=0}^l F_{lmh}(\inc)\sum_{j=-\infty}^{\infty}G_{lhj}(e)
  \sum_{s_2=\,-\,\infty}^{\infty}J_{s_2}(m\,{\cal{A}})\,\sin(v_{lmpq}-\,m\,\theta_{res}~+~\psi_{lm}~-~s_1\,{\cal{M}})
  \;.\;\;\;
 \label{mega2}
 \ea
 Combined with expression (\ref{kilo}), the above two formulae yield:
  \ba
 \nonumber
 \left(\frac{R}{r}\right)^{l+1}\,P_{lm}(\sin\phi)\;\cos\left(\,v^*_{lmpq}-\,m\,(\lambda+\theta^*)~+~\psi_{lm}~-~\epsilon_l\, \right)~=\qquad\qquad\qquad\\
 \nonumber\\
 \nonumber
  \left(\frac{R}{a}\right)^{l+1}\,\sum_{h=0}^{\it l}F_{lmh}(\inc)\sum_{j=-\infty}^{\infty}
  G_{lhj}(e)\sum_{s_1=-\infty}^{\infty}\sum_{s_2=-\infty}^{\infty}J_{s_1}(m\,{\cal{A}})\,J_{s_2}(m\,{\cal{A}})\\
  \nonumber\\
  \;\cos\left(\,
  \left[v_{{\it l}mpq}^*-\,m\theta_{res}^*\,-\,s_1\,{\cal{M}}\right]
  - \left[v_{{\it l}mhj}-m\theta_{res}\,-\,s_2\,{\cal{M}}\right]\,-\,\epsilon_{l}\,\right)
 \,~_{\textstyle{_{\textstyle .}}}
 \label{}
 \ea
 Plugging this result into expression (\ref{bach}), and applying the transformation (\ref{118}) to $\,\theta^*\,$, we end up with
 \ba
 \nonumber
 U(\erbold\,,\;\erbold^{\;*})
 \,=\,-\;\frac{G\,M^*}{a}\;\sum_{l=2}^{\infty}\left(\,\frac{R}{a}\,\right)^{\textstyle{^{2l+1}}}
  \sum_{m=0}^{l}\;\frac{({\it l} - m)!}{({\it l} + m)!}\,
  \left(\,2~-\;\delta_{0m}\,\right)\,
  \sum_{h=0}^{\it l}F_{lmh}(\inc)\sum_{j=-\infty}^{\infty}G_{lhj}(e)\qquad\\
 \nonumber\\
   \sum_{p=0}^l\;F_{lmp}(i^*)\;\sum_{q=\,-\,\infty}^{\infty}\;G_{lpq}(e^*)\;\sum_{s_1=-\infty}^{\infty}\sum_{s_2=-\infty}^{\infty}J_{s_1}(m\,{\cal{A}})\,J_{s_2}(m\,{\cal{A}})\qquad\qquad
   \label{quiddd}
   ~\\
  \nonumber\\
  \nonumber
  ~k_l~\cos\left(\,
  \left[v_{lmpq}^*-\,m\theta_{res}^*\,-\,s_1\,{\cal{M}}\right]
  - \left[v_{{\it l}mhj}-m\theta_{res}\,-\,s_2\,{\cal{M}}\right]\,-\,\epsilon_{l}\,\right)
 \,~_{\textstyle{_{\textstyle ,}}}\,\qquad
 \ea
 where we also inserted phase lags. Be mindful that now the lags and the Love numbers depend on the Fourier modes $\,\beta_{lmpqs}\,$:
 \ba
 k_l\,=\;k_l(\beta_{lmpqs})\;\;,~\quad\epsilon_l\,=\;\epsilon_l(\beta_{lmpqs})\,\;.
 \label{}
 \ea
 At this point, we may recall that $\,\theta_{res}^*\,=\,\theta_{res}\,$, so these two quantities cancel one another.

 For the test particle coinciding with the perturber, expression (\ref{quiddd}) simplifies to
 \ba
 \nonumber
 U(\erbold^{\;*}\,,\;\erbold^{\;*})
 \,=\,-\;\frac{G\,M^*}{a}\;\sum_{l=2}^{\infty}\left(\,\frac{R}{a}\,\right)^{\textstyle{^{2l+1}}}
  \sum_{m=0}^{l}\;\frac{({\it l} - m)!}{({\it l} + m)!}\,
  \left(\,2~-\;\delta_{0m}\,\right)\,
  \sum_{h=0}^{\it l}F_{lmh}(\inc)\sum_{j=-\infty}^{\infty}G_{lhj}(e)\qquad\\
 \nonumber\\
   \sum_{p=0}^l\;F_{lmp}(i)\;\sum_{q=\,-\,\infty}^{\infty}\;G_{lpq}(e)\;\sum_{s_1=-\infty}^{\infty}\sum_{s_2=-\infty}^{\infty}J_{s_1}(m\,{\cal{A}})\,J_{s_2}(m\,{\cal{A}})\qquad\qquad
   \label{quid}
   ~\\
  \nonumber\\
  \nonumber
  ~k_l~\cos\left(\,
  \left[v_{{\it l}mpq}
  \,-\,s_1\,{\cal{M}}\right]
  - \left[v_{{\it l}mhj}
  \,-\,s_2\,{\cal{M}}\right]\,-\,\epsilon_{l}\,\right)
 \,~_{\textstyle{_{\textstyle .}}}\,\qquad\qquad\qquad
 \ea
 This expression comprises oscillating terms and a secular part. The secular part consists of the terms satisfying all three conditions: $\,h=p\,$, $\,j=q\,$, $\,s_1=s_2\;$:
  \ba
 \nonumber
 \langle \;U(\erbold^{\;*}\,,\;\erbold^{\;*})\;\rangle
 \,=\,-\;\frac{G\,M^*}{a}\;\sum_{l=2}^{\infty}\left(\,\frac{R}{a}\,\right)^{\textstyle{^{2l+1}}}
  \sum_{m=0}^{l}\;\frac{({\it l} - m)!}{({\it l} + m)!}\,
  \left(\,2~-\;\delta_{0m}\,\right)\,\qquad\qquad\qquad\\
 \label{buck}\\
 \nonumber
   \sum_{p=0}^l\;F^{\,2}_{lmp}(i)\;\sum_{q=\,-\,\infty}^{\infty}\;G^{\,2}_{lpq}(e)\;\sum_{s=-\infty}^{\infty}J^2_{s}(m\,{\cal{A}})
     \;k_l(\beta_{lmpqs})~\cos\epsilon_l(\beta_{lmpqs})
 \,~_{\textstyle{_{\textstyle .}}}
 \ea
 A similar derivation for the polar torque gives us:
 \ba
 \nonumber
 {\cal{T}}_{polar}^{\rm{^{\,(TIDE)}}}\;=\qquad\qquad\qquad\qquad\qquad\qquad\qquad\qquad\qquad\qquad\qquad\qquad\qquad\qquad\qquad\qquad\qquad\qquad\quad\\
  \nonumber
 -\;\frac{G\,{M^*}^{\,2}}{a}\;\sum_{l=2}^{\infty}\left(\,\frac{R}{a}\,\right)^{\textstyle{^{2l+1}}}
  \sum_{m=1}^{l}\;2\;m\;\frac{(l - m)!}{({\it l} + m)!}
    \sum_{h=0}^{\it l}F_{lmh}(\inc)\sum_{j=-\infty}^{\infty}G_{lhj}(e)\sum_{p=0}^l F_{lmp}(i)\;\sum_{q=\,-\,\infty}^{\infty}G_{lpq}(e)\qquad
    \label{2222}\\
 \nonumber\\
   \sum_{s_1=-\infty}^{\infty}\sum_{s_2=-\infty}^{\infty}J_{s_1}(m\,{\cal{A}})\,J_{s_2}(m\,{\cal{A}})
     ~k_l~\sin\left(v_{lmpq}\,-\,v_{lmhj}\,+\,{\cal{M}}\,(s_2\,-\,s_1)\,-\,\epsilon_{l}\,\right)
 \,~_{\textstyle{_{\textstyle ,}}}\,\qquad\qquad\qquad
 \ea
 its secular part being
 \ba
 \nonumber
 \langle \;{\cal{T}}_{polar}^{\rm{^{\,(TIDE)}}}\;\rangle
 \,=\;\frac{G\,{M^*}^{\,2}}{a}\;\sum_{l=2}^{\infty}\left(\,\frac{R}{a}\,\right)^{\textstyle{^{2l+1}}}
  \sum_{m=1}^{l}\;2\;m\;\frac{({\it l} - m)!}{({\it l} + m)!}\;\,\qquad\qquad\qquad\\
 \label{}\\
 \nonumber
   \sum_{p=0}^l\;F^{\,2}_{lmp}(i)\;\sum_{q=\,-\,\infty}^{\infty}\;G^{\,2}_{lpq}(e)\;\sum_{s=-\infty}^{\infty}J^{\,2}_{s}(m\,{\cal{A}})
     \;k_l(\beta_{lmpqs})~\sin\epsilon_l(\beta_{lmpqs})
 \,~_{\textstyle{_{\textstyle .}}}
 \ea
 In many practical situations, it is convenient to truncate this expression in a manner explained in Makarov et al. (2012) and Noyelles et al. (2014):
 \ba
 \nonumber
 \langle\,{\cal{T}}_{polar}^{\rm{^{\,(TIDE)}}}\,\rangle_{\textstyle{_{\textstyle_{\textstyle{_{l=2}}}}}}~=~~\quad~\quad~~\quad~\quad~\quad~\quad
 ~\quad~\quad~\quad~\quad~\quad~\quad~\quad~\quad~\quad~\quad~\quad~\quad~\quad~\quad~\quad~\quad~\quad~\quad~\quad~\quad\\
 \nonumber\\
 \frac{3}{2}~G\,{M^{\,*}}^{\,2}\,R^5\,a^{-6}\sum_{q=-1}^{7}\,G^{\,2}_{\textstyle{_{\textstyle{_{20\mbox{\it{q}}}}}}}(e)~
 \sum_{s=-s_{\rm{max}}}^{s_{\rm{max}}}J^{\,2}_{s}(m\,{\cal{A}})
 ~k_2(\omega_{\textstyle{_{\textstyle{_{220\mbox{\it{q}}}}}}})~\sin\epsilon_2(\beta_{\textstyle{_{\textstyle{_{220\mbox{\it{q}}s}}}}})
 \,+O(e^8\,\epsilon)+O(\inc^2\,\epsilon)~~.~\quad~\quad~
 \label{overall}
 \label{7c}
 \ea
The number of Bessel functions considered in the above formula (and limited by $\,s_{\rm{max}}\,$) can be determined by trial, depending on the amplitude $\,{\cal{A}}\,$ of the forced libration.



   \section*{\underline{\textbf{\Large{Appendix $\,$D.}}}\vspace{3mm}\\
   ~~~\Large{The permanent-triaxiality-generated torque\\
             expressed through the orbital elements}\label{D}}

 The gravitational potential of an extended body of the mass $\,M\,$ can be expressed as
 \ba
 V(r,\,\phi,\,\lambda)\,=\,-\,\frac{G\,M}{r}\,
 \left[1 - \sum_{l=1}^{\infty} J_l \left(\frac{R}{r}
 \right)^l\,P_l(\sin \phi)
 \right.
 \qquad\qquad\qquad\qquad\qquad\qquad
  \nonumber\\
 \label{409}
 \left.
 + \sum_{l=1}^{\infty}\,\sum^{l}_{m=1} J_{lm}\left(
 \frac{R}{r}\right)^n P_{lm}\left(\sin \phi \right)\;\cos m\left(\lambda
 -\lambda_{lm} \right)\right]
 \;.\;\;
 \ea
 Here $\,\lambda\,$ is the geographic longitude measured eastward from the major axis of the elliptical equatorial cross section that goes through the centre of mass of the planet
 (not through the centre of its figure). The angle $\,\phi\,$ is the latitude. If reckoned from the ascending node, it obeys
 \ba
 \sin \phi\;=\;\sin \inc \; \sin (f\;+\;\omega)\;\;,
 \label{410}
 \ea
 $f\;$ being the true anomaly of the perturber.
 The distance from the origin to the perturber is
 \ba
 r\;=\;a\;\frac{1\;-\;e^2}{1\;+\;e\;\cos f}\;\;\;,
 \label{411}
 \ea
 with $\,a\,,\;e\,,\;\inc\,,\;\omega\,$ being the Keplerian elements of the perturber's apparent orbit~---~the semimajor axis, the eccentricity, the inclination,
 and the argument of the periapse. The quantity $\,R\,$ has dimensions of length. When the extended body has a near-spherical shape, $\,R\,$ acquires the meaning of its equatorial radius.
 The sign convention is chosen as in the physical literature -- so that, for a unit mass, $\;{{\ddoterbold}}\,=\;-\;\nabla V\;$.

 In different notation, the potential may be written also as
 \ba
 V(r,\phi,\lambda)=\;-\;\frac{GM}{r}\left[1-\sum_{l=1}^{\infty}J_l\left(\frac{R}{
 r}\right)^{l} P_l(\sin\phi)
 \right.
 \qquad\qquad\qquad\qquad\qquad\qquad\quad
 \nonumber\\
 ~ \label{412}\\
 \left.
 +\sum_{l=1}^{\infty}\sum_{m=1}^{l}\left(\frac{R}{r}
 \right)^l P_{lm}(\sin\phi)\left[C_{lm} \cos m\lambda +S_{lm} \sin m\lambda\right]
 \right]
 \nonumber
 \ea
or, equivalently, as
 \ba
 V(r,\,\phi,\,\lambda)\,=\,-\,\frac{G\,M}{r}\;\sum_{l=0}^{\infty}\;\sum_{m=0}^{l}\;\left(\,
 \frac{R}{r}\,\right)^l\;P_{lm}(\sin\phi)\;\left[\,C_{lm}\;\cos m\lambda\;+\;
 S_{lm}\;\sin m\lambda\,\right] \;\;,\;\;~~~~~~~~~
 \label{413}
 \ea
where, for any $\;l\;$,
 \ba
 P_{l}(\sin\phi)\,=\,P_{l0}(\sin\phi)~.~~~~
 \label{414}
 \ea
 For $~l\,=~\,2\,,\,.\,.\,.\,,\,\infty~$ and $~m\,=\,1\,,\,.\,.\,.\,,\,l~\,$:
 \ba
 C_{lm}\;\equiv\;J_{lm}\;\cos m\lambda_{lm} ~~~,~~~~~
 S_{lm}\;\equiv\;J_{lm}\;\sin m\lambda_{lm} ~~~,~~~
 \label{415}
 \ea
 while for $~l\,=~\,2\,,\,.\,.\,.\,,\,\infty~$ and $~m\,=\,0~$ we have
 \ba
 C_{{{l}}0}\;=\;-\;J_{{{l}}0}\,\equiv\,-\;J_l~~~.
 \label{416}
 \ea
 For $~l\,=~\,0~$, we get:
 \ba
 C_{00}\,\equiv\,J_{00}\,\equiv\,1 ~~~,~~~P_{0}(\sin\phi)\,=\,P_{00}(\sin\phi)\,\equiv\,1~~~.~
 \label{417}
 \ea
 The $\;l\,=\,1\;$ terms need some attention. For a general choice of the origin,
 \ba
 C_{11}\;=\;\frac{X_{cm}}{R}~~~,~~~~~
 S_{11}\;=\;\frac{Y_{cm}}{R}~~~,~~~~~
 C_{10}\,\equiv\,-\,J_{10}\,\equiv\,-\;J_1\;=\;\frac{Z_{cm}}{R}~~~,~~~~~
 \label{418}
 \ea
 $X_{cm}\,,~Y_{cm}\,,~Z_{cm}~$ being the Cartesian coordinates of the centre of mass (Hobson
 1965). The coefficient $\;J_{11}\;$ is then related to $\;C_{11}\;$ and $\;S_{11}\;$ through
 (\ref{415}). If however we choose to place the origin in the centre of mass, we obtain:
 \ba
 C_{10}\,\equiv\,-\,J_{10}\,\equiv\,-\;J_1\;=\;0~~~,~~~~\;C_{11}\;=\;0~~~,~~~~S_{11}\;=\;0
 ~~~,~~~~~J_{11}\;=\;0~~~,~~~~~
 \label{419}
 \ea
 which nullifies all the $\,{\it{l}}=1\,$ terms in (\ref{409}), (\ref{412}), and (\ref{413}). The terms containing $\;C_{{{l}}0}=\,-\,J_{{{l}}0}\,\equiv\,-\,J_l\;$
 are called zonal; the terms for which $l\ne j \ne 0$ are called tesseral, while those with $l=j$ are called sectorial.

 Combining expression (\ref{413}) with Kaula's trigonometric formula (\ref{kaula}), we arrive at:
 \ba
 V(r,\,\phi,\,\lambda)\,=\,-\,\frac{G\,M}{a}\;\sum_{l=0}^{\infty}\;\sum_{m=0}^{l}\;\left(\,
 \frac{R}{a}\,\right)^l\;\sum_{p=0}^{\infty} F_{lmp}(i)\sum_{q=\,-\,\infty}^{\infty}G_{lpq}(e)\qquad\qquad\qquad\qquad\qquad\qquad\qquad
 \nonumber\\
 \label{}\\
 \left[\,C_{lm}\;
  \left\{
  \begin{array}{c}
  \cos \left(\,
  v_{{\it l}mpq}\,-\,m\,\theta\,\right)\,\\
  \sin \left(\,
  v_{{\it l}mpq}\,-\,m\,\theta\,\right)\,
  \end{array}
  \right\}^{{\it l}\,-\,m\;\;
  \mbox{\small even}}_{{\it l}\,-\,m\;\;\mbox{\small odd}}
 \;+\;
 S_{lm}\;
 \left\{
  \begin{array}{c}
  \left.\;\;\right.\sin \left(\,
  v_{{\it l}mpq}\,-\,m\,\theta\,\right)\,\\
  -\;\cos \left(\,
  v_{{\it l}mpq}\,-\,m\,\theta\,\right)\,
  \end{array}
  \right\}^{{\it l}\,-\,m\;\;
  \mbox{\small even}}_{{\it l}\,-\,m\;\;\mbox{\small odd}}
 \,\right] \;\;{_{\textstyle{_{\textstyle{_{\textstyle{_{\textstyle{_{\textstyle{_{\textstyle.}}}}}}}}}}}}~~~~
 \nonumber
 \ea
 Multiplying this by the mass $\,M^*\,$ of the point perturber, we obtain the potential energy of interaction of the two bodies.
 The polar torque acting on the extended body is the negative of the derivative of the energy with respect to the rotational angle $\,\theta\,$:
 \bs
 \ba
 {\cal{T}}_{polar}^{\rm{^{\;(TRI)}}}=\;-\;\frac{G\,M\,M^*}{a}\;\sum_{l=0}^{\infty}\;\sum_{m=1}^{l}\;\left(\,
 \frac{R}{a}\,\right)^l m\;\sum_{p=0}^{\infty} F_{lmp}(i)\sum_{q=\,-\,\infty}^{\infty}G_{lpq}(e)\qquad\qquad\qquad\qquad\qquad\qquad
 \nonumber\\
 \label{}\\
 \left[\,C_{lm}\;
  \left\{
  \begin{array}{c}
  -\;\sin \left(\,
  v_{{\it l}mpq}\,-\,m\,\theta\,\right)\,\\
  \left.\;\;\right.\cos \left(\,
  v_{{\it l}mpq}\,-\,m\,\theta\,\right)\,
  \end{array}
  \right\}^{{\it l}\,-\,m\;\;
  \mbox{\small even}}_{{\it l}\,-\,m\;\;\mbox{\small odd}}
 \;+\;
 S_{lm}\;
 \left\{
  \begin{array}{c}
  \cos \left(\,
  v_{{\it l}mpq}\,-\,m\,\theta\,\right)\,\\
  \sin \left(\,
  v_{{\it l}mpq}\,-\,m\,\theta\,\right)\,
  \end{array}
  \right\}^{{\it l}\,-\,m\;\;
  \mbox{\small even}}_{{\it l}\,-\,m\;\;\mbox{\small odd}}
 \,\right] \;\;{_{\textstyle{_{\textstyle{_{\textstyle{_{\textstyle{_{\textstyle{_{\textstyle .}}}}}}}}}}}}~~~~
 \nonumber
 \label{}
 \ea
 For a homogeneous triaxial ellipsoid, vanish all the coefficients $\,S_{lm}\,$, as well as those $\,C_{lm}\,$ for which at least one index is odd.$\,$\footnote{~To understand why the coefficients $\,S_{lm}\,$ vanish for an ellipsoid, change $\,\lambda\,$ to $\;-\,\lambda\,$ in expression
 (\ref{412}) or (\ref{413}). The symmetry then necessitates $\,S_{lm}\,=\,0\,$, under the condition that $\,\lambda\,$ is reckoned $\,${\it{from a principal axis.}}
 Under the same condition, should vanish all $\,C_{lm}\,$ for an odd $\,m\,$, because for an elliptic body the said expressions must stay unaltered under the change of $\,\lambda\,$ to $\,\lambda+\pi\,$. Likewise, the potential of an ellipsoid, (\ref{412}) or (\ref{413}), should not change under the change of $\,\phi\,$ to $\,\phi+\pi\,$; hence the vanishing of all $\,C_{lm}\,$ for an odd $\,l\,$~---~recall that $\,P_{lm}(-x)=(-1)^lP_{lm}(x)\,\,$.  $\,$Also mind that the node $\,\Omega\,$ and the rotation angle $\,\theta\,$ must be reckoned from the same fiducial direction (usually, the vernal equinox).
 \label{important}}
 The only survivors are $\,C_{lm}\,$ with both $\,l\,$ and $\,m\,$ even. Thence,
 \ba
 ^{\textstyle{^{\rm{(ellipsoid)}}}}{\cal{T}}_{polar}^{\rm{^{\;(TRI)}}}= \qquad\qquad\qquad\qquad\qquad\qquad\qquad\qquad\qquad\qquad\qquad\qquad\qquad\qquad\qquad\qquad\qquad\qquad
 \nonumber\\
 \label{ellipsoid}\\
 \nonumber
  \frac{G\,M\,M^*}{a}\;\sum_{\stackrel{l=0}{{(l\;{\rm{even}})}}}^{\infty}\sum_{\stackrel{m=2}{{(m\;{\rm{even}})}}}^{l}\left(
 \frac{R}{a}\right)^l m\;C_{lm}\,\sum_{p=0}^{\infty} F_{lmp}(i)\sum_{q=\,-\,\infty}^{\infty}G_{lpq}(e)
 \;
  \sin \left(\,v_{{\it l}mpq}\,-\,m\,\theta\,\right)
 \,~.~~
 \ea
  ~\\
 In the limit of vanishing $\,i\,$, only those $\,F_{lmp}\,$ stay nonzero for which, according to Gooding \& Wagner (2008, Section 9.2), the indices obey $\,p\,=\,(l-m)/2\,$. This entails:
 \ba
 ^{\textstyle{^{\rm{(ellipsoid)}}}}{\cal{T}}_{polar}^{\rm{^{\;(TRI)}}}= \qquad\qquad\qquad\qquad\qquad\qquad\qquad\qquad\qquad\qquad\qquad\qquad\qquad\qquad\qquad\qquad\qquad\qquad
 \nonumber\\
 \label{}\\
 \nonumber
  \frac{G\,M\,M^*}{a}\;\sum_{\stackrel{\textstyle{_{l=0}}}{{(l\;{\rm{even}})}}}^{\infty}\sum_{\stackrel{\textstyle{_{m=2}}}{{(m\;{\rm{even}})}}}^{l}\left(
 \frac{R}{a}\right)^l m\;C_{lm}\, F_{lmp}(0)\sum_{q=\,-\,\infty}^{\infty}G_{lpq}(e)
 \;
  \sin \left(\,v_{lmpq}\,-\,m\,\theta\,\right)_{\textstyle{_{\textstyle{_{\rm{p=(l-m)/2}}}}}}+\;O(i)\;.\qquad
 \ea
 Specifically,
 \ba
 F_{220}(\inc)=3+O(\inc^2)~\,,~~~\,F_{210}(\inc)=\frac{3}{2}\,\sin\inc +
 O(\inc^2)~\,,~~~\,F_{211}(\inc)=\,-\,\frac{3}{2}\,\sin\inc +O(\inc^2)~\,,~\quad\,\quad
 \nonumber
 \ea
 all the other $F_{2mp}(\inc)$ being of order $O(\inc^2)$ or higher. Thence, in the leading order of $\,i\,$, the quadrupole ($\,l=2\,$) part of the torque is:
 \ba
 _{\textstyle{_{\rm{(quadrupole)}}}}^{\textstyle{^{\rm{(ellipsoid)}}}}{\cal{T}}_{polar}^{\rm{^{\;(TRI)}}}=
 \;6\;\frac{G\,M\,M^*}{a}\;\left(\frac{R}{a}\right)^2 C_{22}\,\sum_{q=\,-\,\infty}^{\infty}G_{20q}(e)
 \;\sin \left(\,v_{220q}\,-\,m\,\theta\,\right)\;+\;O(i)\qquad\;
 \label{}\\
 \nonumber\\
 =\;\frac{3}{2}\;\frac{G\,M^*}{a^3}\;(B\,-\,A)\,\sum_{q=\,-\,\infty}^{\infty}G_{20q}(e)
 \;\sin 2\left[\,\left(1\,+\,\frac{\textstyle q}{\textstyle 2}\right)\,{\cal{M}}\,+\,\omega\,+\,\Omega\,-\,\theta\,\right]\;+\;O(i)
 \;\,,\qquad
 \label{final}
 \ea
 \label{survivor}
 \es
 where we used the relation
 \ba
 C_{22}\,=\,\frac{B\,-\,A}{4\,M\,R^{\,2}}\,\;.
 \label{}
 \ea

   \section*{\underline{\textbf{\Large{Appendix $\,$E.}}}\vspace{3mm}\\
   ~~~\Large{Dissipation rate due to the tidal forces in a librating body}\label{E}}

   \subsection*{E.1~~~Generalities}

  The overall tide-raising potential $\,W\,$ is generated by the perturber and by the second-degree part of the centrifugal force. Both these inputs cause deformation of shape and, consequently, an additional potential $\,U\,$ of the tidally deformed body. While a degree-$l$ part of external perturbing potential, $\,W_l\,$, generates a degree-$l$ input in the additional tidal potential,
  the quadrupole part $\,W_2^{(cent)}\,$ of the centrifugal potential generates a quadrupole input $\,U_2^{(cent)}\,$ (see, e.g., Efroimsky 2017):
 \ba
 W\;=\;W_2^{(cent)}\,+\;\sum_{l=2}^{\infty}W_l\quad,\qquad U\;=\;U_2^{(cent)}\,+\,\sum_{l=2}^{\infty}U_l\,\;.
 \label{equ}
 \ea
 For each degree $\,l\,$, both $\,W_l\,$ and $\,U_l\,$ are proportional to the Legendre polynomial $\,P_l(\cos\gamma)\,$, where $\,\gamma\,$ is the angular separation between the planetocentric radii aimed at the perturber and at a point wherein a potential is measured.

 For a body that is spin-synchronised (i.e., showing the same side to the perturber), the expression for the energy damping rate was derived by Peale \& Cassen (1978). For a body rotating outside the 1:1 resonance, that theory was generalised by Efroimsky \& Makarov (2014). In the case of libration, further generalisation is needed, because in this situation an extra ``quantum number'' is present in the spectrum of tidal modes.

 With angular brackets denoting time averaging, a general expression for the damped power in an arbitrary spin state can be written as
 (Efroimsky \& Makarov 2014, eqn 61):
 \begin{equation}
 \langle \,P\,\rangle_{\textstyle{_{\rm{tide}}}}\,=\;\frac{1}{4\,\pi\,G\,R}\;\sum_{l\,=\,2}^{\infty}\,(2l+1) \int \langle \,W_l \;\dot{U}_l\,\rangle\;dS\;\,.
 \label{equa}
 \end{equation}
 The expression (\ref{equa}) is general in that it stays valid also when the quadrupole parts $\,W_2\,$ and $\,U_2\,$ include the quadrupole inputs from the centrifugal potential, as in equation (\ref{equ}). However, hereafter we shall proceed with only the gravitational tide taken into account. The reason for this is explained below.

 \subsection*{E.2~~~Difficulty}

 Since the centrifugal force is quadratic in the angular velocity, it can be demonstrated that a frequency $\,\chi_j\,=\,j\,n\,$ in the libration spectrum generates a frequency
 $\,2\,\chi_j\,=\,2\,j\,n\,$ in the alternating part of the centrifugal force. Each of these coincides with the physical frequency produced by some of the gravitational tidal modes: $\;\,2\,\chi_j\,=\,|\,\beta_{lmpqs}\,|\,$,  which is (in neglect of $\,\dot\omega\,$) equivalent to $\;\,2\,j\,=\,|\,l\,-\,2\,p\,-\,m\,z\,+\,q\,-\,s\,|\,$.

  Hence the method of finding the total damping rate must be as follows. In each spin-orbit resonance $\,z\,$, we must write down the terms of the gravitational tidal potential, that we choose to preserve. Along with that, we must write down the terms of the quadrupole potential due to the centrifugal force, that we keep. Then, for each even frequency $\,2\,\chi_j\,=\,2\,j\,n\,$, we should sum up the corresponding terms from both these groups, and should calculate the resulting power at each frequency separately.
  At odd frequencies, only the gravitational tide will enter the calculation, while at the even frequencies $\,2\,\chi_j\,=\,2\,j\,n\,$ both the gravitational and centrifugal parts will come into play.

  We may avoid this cumbersome calculation when one of these two effects is negligible compared to another. As demonstrated in Efroimsky (2017), in the case of small-magnitude libration the input from the centripetal force is small and may be omitted.

  \subsection*{E.3~~~Tidal heating due to the gravitational tide only.\\
  The case of forced libration\label{forced}}

 As we saw in Section \ref{gen}, the components of the potentials due to the gravitational tide are given by the following expansions:
  \begin{equation}
 W_l\,=\;\sum_{m=0}^l \sum_{p=0}^l \sum_{q=-\infty}^{\infty} \sum_{s=-\infty}^{\infty} C_{lmpqs} \;\cos(B_{lmpq}\,-\,s\,{\cal{M}})
 \label{gerd}
 \end{equation}
 \begin{equation}
 U_l\,= \sum_{m=0}^l \sum_{p=0}^l \sum_{q=-\infty}^{\infty} \sum_{s=-\infty}^{\infty} C_{lmpqs}\;k_l(\beta_{lmpqs})\;\cos(B_{lmpq}\,-\,s\,{\cal{M}}\,-\,\epsilon_l(\beta_{lmpqs})\,)
 \label{gerdt}
 \end{equation}
 where $\,B_{lmpq}\,$ are some linear combinations of the orbital elements.

 The quantities $\,\beta_{lmpqs}\,$, $\,\epsilon_l(\beta_{lmpqs})\,$, and $\,k_l(\beta_{lmpqs})\,$ are the libration-caused tidal modes and the corresponding phase lags and Love numbers. The expression for the tidal mode $\,\beta_{lmpqs}\,$ is given above by formula (\ref{119}).

 The coefficients $\,C_{lmpqs}\,$ entering expression (\ref{gerdt}) depend on the amplitude $\,{\cal{A}}\,$ of the forced libration:
\begin{equation}
C_{lmpqs}\,=\,-\,\frac{GM^*}{a} \bigg( \frac{R}{a} \bigg)^l \frac{(l-m)!}{(l+m)!} (2-\delta_{0m}) P_{lm}(\sin \phi) F_{lmp} (i) G_{lpq}(e) J_s (m {\cal{A}})\,\;.
\end{equation}
  This expression was derived for the case when the longitudinal libration contains only the principal mode $\,\theta\,=\,{\cal{A}}\,\sin{\cal{M}}\,=\,{\cal{A}}\,\sin nt\,$, with $\,n\,\equiv\,\dot{\cal{M}}\,$ being the anomalistic mean motion. Our formalism, however, can be generalised to the case of several frequencies, along the lines explained in Appendix A.2.

 The product of the two above potentials reads as
 \begin{equation}
 \begin{split}
 W_l\,\dot{U}_l =\,
 -\,\frac{1}{2} \sum_{m^{\,\prime}=0}^l \sum_{p^{\,\prime}=0}^l \sum_{q^{\,\prime}=-\infty}^{\infty} \sum_{s^{\,\prime}=-\infty}^{\infty}
\sum_{m=0}^l \sum_{p=0}^l \sum_{q=-\infty}^{\infty} \sum_{s=-\infty}^{\infty}
C_{lm^{\,\prime}p^{\,\prime}q^{\,\prime}s^{\,\prime}} \, C_{lmpqs} \, \beta_{lmpqs} \, k_l(\beta_{lmpqs})  \\
\times \sin([l-2p^{\,\prime}-m^{\,\prime}z + l-2p-mz] \omega + [l-2p^{\,\prime}+q^{\,\prime}-m^{\,\prime}z - s^{\,\prime} + l-2p+q-mz - s] {\cal{M}} \\
+ [m^{\,\prime}-m^{\,\prime}z + m-mz] \Omega + [-m^{\,\prime} - m] \lambda +  \psi_{lm^{\,\prime}} + \psi_{lm} - \epsilon_l(\beta_{lmpqs}))  \\
+ \frac{1}{2} \sum_{m^{\,\prime}=0}^l \sum_{p^{\,\prime}=0}^l \sum_{q^{\,\prime}=-\infty}^{\infty} \sum_{s^{\,\prime}=-\infty}^{\infty}
\sum_{m=0}^l \sum_{p=0}^l \sum_{q=-\infty}^{\infty} \sum_{s=-\infty}^{\infty}
C_{lm^{\,\prime}p^{\,\prime}q^{\,\prime}s^{\,\prime}} \, C_{lmpqs} \, \beta_{lmpqs} \, k_l(\beta_{lmpqs})  \\
\times \sin([l-2p^{\,\prime}-m^{\,\prime}z - l+2p+mz] \omega + [l-2p^{\,\prime}+q^{\,\prime}-m^{\,\prime}z - s^{\,\prime} - l+2p-q+mz + s] {\cal{M}} \\
+ [m^{\,\prime}-m^{\,\prime}z - m+mz] \Omega + [-m^{\,\prime} + m] \lambda+ \psi_{lm^{\,\prime}} - \psi_{lm}  + \epsilon_l(\beta_{lmpqs}))\,\;.
\end{split}
\end{equation}
 In simplifying this expression, the first step can be made by considering the outcome of the integration over the longitude $\,\lambda\,$. In the first sum above, only the terms with $m^{\,\prime} = m = 0$ will survive averaging, while in the second sum the only survivors will be the terms with $m=m^{\,\prime}$. In the first sum, the presence of the resulting phase $\,2 \psi_{l0}\,$ is equivalent to multiplying that sum with $\,(-1)^l\,$. Then we can rewrite the above expression as
 \begin{equation}
 \begin{split}
W_l \dot{U}_l =
- \frac{(-1)^l}{2}  \sum_{p^{\,\prime}=0}^l \sum_{q^{\,\prime}=-\infty}^{\infty} \sum_{s^{\,\prime}=-\infty}^{\infty}
\sum_{p=0}^l \sum_{q=-\infty}^{\infty} \sum_{s=-\infty}^{\infty}
C_{l0p^{\,\prime}q^{\,\prime}s^{\,\prime}} \, C_{l0pqs} \, \beta_{l0pqs} \, k_l(\beta_{l0pqs})  \\
\times \sin([2l-2p^{\,\prime} -2p] \omega + [2l-2p^{\,\prime}+q^{\,\prime} - s^{\,\prime} -2p+q - s] {\cal{M}}    - \epsilon_l(\beta_{l0pqs}))  \\
 + \frac{1}{2} \sum_{p^{\,\prime}=0}^l \sum_{q^{\,\prime}=-\infty}^{\infty} \sum_{s^{\,\prime}=-\infty}^{\infty}
\sum_{m=0}^l \sum_{p=0}^l \sum_{q=-\infty}^{\infty} \sum_{s=-\infty}^{\infty}
C_{lmp^{\,\prime}q^{\,\prime}s^{\,\prime}} \, C_{lmpqs} \, \beta_{lmpqs} \,k_l(\beta_{lmpqs})  \\
\times \sin([-2p^{\,\prime} +2p] \omega + [-2p^{\,\prime}+q^{\,\prime} - s^{\,\prime} +2p-q + s] {\cal{M}} + \epsilon_l(\beta_{lmpqs})) \\
+...
\label{lies}
 \end{split}
 \end{equation}
 where the ellipsis denotes the terms that will vanish after averaging over the longitude $\,\lambda\,$.

 As the second step, we average over the motion of the pericentre $\,\omega\,$ and over the mean longitude ${\cal{M}}$. In the first sum, only the terms with $\,p^{\,\prime} = l-p\,$ and $\,s^{\,\prime} = q^{\,\prime} + q - s\,$ survive this procedure; in the second sum, only those with $\,p^{\,\prime} = p\,$ and $\,s^{\,\prime} = q^{\,\prime}-q+s\,$. Thus we arrive at
  \begin{equation}
\begin{split}
\langle W_l \dot{U}_l \rangle =
 \frac{(-1)^l}{2}  \sum_{q^{\,\prime}=-\infty}^{\infty} \sum_{p=0}^l \sum_{q=-\infty}^{\infty} \sum_{s=-\infty}^{\infty}
C_{l0(l-p)q^{\,\prime}(q^{\,\prime}+q-s)} \, C_{l0pqs} \, \beta_{l0pqs} \, k_l(\beta_{l0pqs})  \sin\epsilon_l(\beta_{l0pqs})  \\
 + \frac{1}{2}  \sum_{q^{\,\prime}=-\infty}^{\infty} \sum_{m=0}^l \sum_{p=0}^l \sum_{q=-\infty}^{\infty} \sum_{s=-\infty}^{\infty}
C_{lmpq^{\,\prime}(q^{\,\prime}-q+s)} \, C_{lmpqs} \, \beta_{lmpqs} \,k_l(\beta_{lmpqs}) \sin\epsilon_l(\beta_{lmpqs}) \\
+... \\
\label{true}
\end{split}
\end{equation}
 Given that $\,J_n (0) = 0\,$ for $\,n\neq 0\,$, and  $\,J_0 (0) = 1\,$, the first sum can be written as
\begin{equation}
\begin{split}
\bigg( \frac{G M^*}{a} \bigg)^2 \bigg( \frac{R}{a} \bigg)^{2l}  \frac{(-1)^l}{2} \sum_{q^{\,\prime}=-\infty}^{\infty} \sum_{p=0}^l \sum_{q=-\infty}^{\infty} \sum_{s=-\infty}^{\infty} P_{l0}(\sin \phi)^2 F_{l0(l-p)} (i) F_{l0p} \, G_{l(l-p)q^{\,\prime}}(e) \, G_{lpq}(e) \\
J_{(q^{\,\prime}+q-s)} (0) J_s (0) \, \beta_{l0pqs} \, k_l(\beta_{l0pqs})  \sin\epsilon_l(\beta_{l0pqs}) \\
=
\bigg( \frac{G M^*}{a} \bigg)^2 \bigg( \frac{R}{a} \bigg)^{2l}  \frac{1}{2}  \sum_{p=0}^l \sum_{q=-\infty}^{\infty} P_{l0}(\sin \phi)^2 F_{l0p}^2 \, G^2_{lpq}(e)  \, \beta_{l0pq0} \, k_l(\beta_{l0pq0})  \sin\epsilon_l(\beta_{l0pq0})\,\;. \\
\end{split}
\end{equation}
 The second sum can be shaped as
\begin{equation}
\begin{split}
\bigg( \frac{G M^*}{a} \bigg)^2 \bigg( \frac{R}{a} \bigg)^{2l} \frac{1}{2}  \sum_{q^{\,\prime}=-\infty}^{\infty} \sum_{m=0}^l \sum_{p=0}^l \sum_{q=-\infty}^{\infty} \sum_{s=-\infty}^{\infty}
\bigg[ \frac{(l-m)!}{(l+m)!} (2-\delta_{0m}) \bigg]^2\, P_{lm}(\sin \phi)^2 \,  F_{lmp}^2 (i)  \, G_{lpq^{\,\prime}}(e) \, G_{lpq}(e) \, \\
J_{(q^{\,\prime}-q+s)} (m {\cal{A}}) J_s (m {\cal{A}}) \,  \beta_{lmpqs} \,k_l(\beta_{lmpqs}) \sin\epsilon_l(\beta_{lmpqs}) \\
=
\bigg( \frac{G M^*}{a} \bigg)^2 \bigg( \frac{R}{a} \bigg)^{2l} \frac{1}{2}   \sum_{p=0}^l \sum_{q=-\infty}^{\infty}
 P_{l0}(\sin \phi)^2 \,  F_{l0p}^2 (i)  \, G^2_{lpq}(e) \,  \beta_{lmpq0} \,k_l(\beta_{lmpq0}) \sin\epsilon_l(\beta_{lmpq0}) \\
+
\bigg( \frac{G M^*}{a} \bigg)^2 \bigg( \frac{R}{a} \bigg)^{2l} \frac{1}{2}  \sum_{q^{\,\prime}=-\infty}^{\infty} \sum_{m=1}^l \sum_{p=0}^l \sum_{q=-\infty}^{\infty} \sum_{s=-\infty}^{\infty}
\bigg[ \frac{(l-m)!}{(l+m)!} (2-\delta_{0m}) \bigg]^2\, P_{lm}(\sin \phi)^2 \,  F_{lmp}^2 (i)  \, G_{lpq^{\,\prime}}(e) \, G_{lpq}(e) \, \\
J_{(q^{\,\prime}-q+s)} (m {\cal{A}}) J_s (m {\cal{A}}) \,  \beta_{lmpqs} \,k_l(\beta_{lmpqs}) \sin\epsilon_l(\beta_{lmpqs})\,\;.\\
\end{split}
\end{equation}
 The integration over the volume can be performed with aid of the formula
\begin{equation}
\int [P_{lm} (\sin \phi)]^2 dS = 2 \pi R^{\,2} \frac{2}{2l+1}    \frac{(l+m)!}{(l-m)!}\,\;.
\end{equation}
 Gathering all our formulae together, we obtain:
\begin{equation}
\begin{split}
\langle P \rangle
& = \frac{1}{4 \pi G R} \sum_{l=2}^{\infty} (2l+1) \bigg( \frac{G M^*}{a} \bigg)^2 \bigg( \frac{R}{a} \bigg)^{2l} \frac{1}{2}  \sum_{q^{\,\prime}=-\infty}^{\infty} \sum_{m=0}^l \sum_{p=0}^l \sum_{q=-\infty}^{\infty} \sum_{s=-\infty}^{\infty}
(1+ \delta_{0m}) \bigg[ \frac{(l-m)!}{(l+m)!} (2-\delta_{0m}) \bigg]^2\, \\
& 2 \pi R^{\,2} \frac{2}{2l+1}    \frac{(l+m)!}{(l-m)!} \,  F_{lmp}^2 (i)  \, G_{lpq^{\,\prime}}(e) \, G_{lpq}(e) \, J_{(q^{\,\prime}-q+s)} (m {\cal{A}}) J_s (m {\cal{A}}) \,  \beta_{lmpqs} \,k_l(\beta_{lmpqs}) \sin\epsilon_l(\beta_{lmpqs})
\end{split}
\end{equation}
 which, after some algebra, yields:
\begin{equation}
\begin{split}
\langle P \rangle &
=
 \frac{G M^{*\,2}}{a} \sum_{l=2}^{\infty}    \bigg( \frac{R}{a} \bigg)^{2l+1}   \sum_{q^{\,\prime}=-\infty}^{\infty} \sum_{m=0}^l \sum_{p=0}^l \sum_{q=-\infty}^{\infty} \sum_{s=-\infty}^{\infty}
  \frac{(l-m)!}{(l+m)!} (2-\delta_{0m}) \, \\
&     \,  F_{lmp}^{\,2} (i)  \, G_{lpq^{\,\prime}}(e) \, G_{lpq}(e) \, J_{(q^{\,\prime}-q+s)} (m {\cal{A}}) J_s (m {\cal{A}}) \,  \beta_{lmpqs} \,k_l(\beta_{lmpqs}) \sin\epsilon_l(\beta_{lmpqs})\,\;.
\label{terra}
\end{split}
\end{equation}
  In the limit of no libration ($\,{\cal A} = 0\,$), only the terms with $\,s=0\,$ and $\,q^{\,\prime} = q\,$ are left, and we recover the result from Efroimsky \& Makarov (2014, eqn 65).

 For small-amplitude libration ($\,m\,{\cal A}\,\ll\,1\,$), we can use the asymptotic expressions
 \ba
 J_s(m\,{\cal{A}})\,\approx\;
 \left\{
 \begin{array}{ll}
 \frac{\textstyle 1}{\textstyle s!}\;\left(\frac{\textstyle m\,{\cal{A}}}{\textstyle 2}\right)^{\rm s}\quad            & \mbox{for~integer}~s\,>\,0\\
 ~\\
 1\;-\;\left(\frac{\textstyle m\,{\cal{A}}}{\textstyle 2}\right)^2             & \mbox{for~integer}~s\,=\,0\\
 ~\\
 \frac{\textstyle (-1)^{\rm s}}{\textstyle (-s)!}\;\left(\frac{\textstyle m\,{\cal{A}}}{\textstyle 2}\right)^{\rm -s}\quad & \mbox{for~integer}~s\,<\,0\;\;\;.
 \end{array}
 \right.
 \label{asymptotic}
 \ea

 To draw to a close, we would reiterate that the above formulae have been written down for the longitudinal libration approximated with its principal mode $\,\theta\,=\,{\cal{A}}\,\sin{\cal{M}}\,=\,{\cal{A}}\,\sin nt\,$. This machinery, however, can be generalised to the case of several libration frequencies. In that case, as we saw in Appendix A.2, we get only one extra quantum number $\,s\,$, no matter how many frequencies enter the longitudinal libration spectrum.

 \subsection*{E.4~~~Tidal heating due to the gravitational tide only.\\
  The case of free libration\label{fr}}

 Consider sinusoidal free libration at a frequency $\,\chi\,$, with some initial phase $\,\varphi\,$. Our derivation in Appendix A.1 remains in force, except that now, instead of equation (\ref{A6c}), we have:
 \ba
 \cos(\,B_{lmpq}\,-m~{\cal A}~\sin (\chi t + \varphi)\,)~=\,\sum_{s=-\infty}^{\infty}\,J_{s}(m{\cal A})~\cos(B_{lmpq}\,-\,s\,(\chi t + \varphi)\,)\,~.
 \label{}
 \ea
 Consequently, the true Fourier tidal modes will now be given not by equation (\ref{119}), but by
 \ba
 \nonumber
 \beta_{lmpqs}\,=~{\bf{\dot{\rm{\mbox{$B$}}}}}_{lmpq}\,-~s\,\chi &=&(l\,-\,2\,p\,-\,m\,z\,')\,\stackrel{\bf\centerdot}{\omega}\,+~(l\,-\,2\,p\,-\,
 m\,z\,+\,q)\,\stackrel{\bf\centerdot}{\cal M\,}\;-\;s\;\chi\\
 \label{11119}\\
 \nonumber
 &\approx &~(l\,-\,2\,p\,-\,
 m\,z\,+\,q)\,\stackrel{\bf\centerdot}{\cal M\,}\;-\;s\;\chi\;\,.
 \ea

 To calculate the dissipation rate, we have to repeat the calculation similar to the one carried out above in Appendix E.3. This time, however, instead of formulae (\ref{gerd}) and (\ref{gerdt}), we have to use, correspondingly,
  \begin{equation}
 W_l\,=\;\sum_{m=0}^l \sum_{p=0}^l \sum_{q=-\infty}^{\infty} \sum_{s=-\infty}^{\infty} C_{lmpqs} \;\cos(B_{lmpq}\,-\,s\,(\chi t + \varphi)\,)
 \end{equation}
 and
 \begin{equation}
 U_l\,= \sum_{m=0}^l \sum_{p=0}^l \sum_{q=-\infty}^{\infty} \sum_{s=-\infty}^{\infty} C_{lmpqs}\;k_l(\beta_{lmpqs})\;\cos(B_{lmpq}\,-\,s\,(\chi t + \varphi)\,-\,\epsilon_l(\beta_{lmpqs})\,)\,\;,
 \label{}
 \end{equation}
 with $\,\beta_{lmpqs}\,$ now given by expression (\ref{11119}).

 Calculating the power, we shall obtain an equation similar to (\ref{lies}), except for one difference: the new equation will contain $\,s\,(\chi t + \varphi)\,$ and $\,s^{\,\prime}\,(\chi t + \varphi)\,$ instead of $\,s{\cal{M}}\,$ and $\,s^{\,\prime}{\cal{M}}\,$.

 While the principal frequency of the forced libration coincided with the mean motion $\,n\,$, the frequency of free libration $\,\chi\,$ is different from it. Accordingly, time averaging of the power will now contain not two but three independent averagings. Indeed, while in Appendix E.3 we averaged over the mean motion and the apsidal precession, now we should average, independently, also over the period of free libration. This will render us an expression similar to (\ref{true}), with two important alterations.
 In the first line of that expression, we now shall have only the terms with $\,q^{\,\prime}=-q\,$, while in the second line only the terms with $\,q^{\,\prime}=q\,$ will survive. In the end, we shall arrive at an expression similar to (\ref{terra}), though with $\,q^{\,\prime}=q\;$:
 \begin{equation}
\begin{split}
\langle P \rangle &
=
 \frac{G M^{*\,2}}{a} \sum_{l=2}^{\infty}    \bigg( \frac{R}{a} \bigg)^{2l+1}    \sum_{m=0}^l \sum_{p=0}^l \sum_{q=-\infty}^{\infty} \sum_{s=-\infty}^{\infty}
  \frac{(l-m)!}{(l+m)!} (2-\delta_{0m}) \, \\~\\
&     \,  F_{lmp}^{\,2} (i)  \, \, G_{lpq}^2(e) \, J_s^2 (m {\cal{A}}) \,  \beta_{lmpqs} \,k_l(\beta_{lmpqs}) \sin\epsilon_l(\beta_{lmpqs})\,\;.
\label{dud}
\end{split}
\end{equation}

 \subsection*{E.5~~~The case of free and forced librations combined}

 Insertion of expression (\ref{labour}) into the formula for the tidal dissipation rate leads to a calculation much more laborious than that presented in Appendix E.3 for forced libration or in Appendix E.4 for free libration. For example, the analogue to equation (\ref{lies}) will read as
  \begin{equation}
 \begin{split}
W_l \dot{U}_l =
- \frac{(-1)^l}{2}  \sum_{p^{\,\prime}=0}^l \sum_{q^{\,\prime}=-\infty}^{\infty} \sum_{s_1^{\,\prime}=-\infty}^{\infty}\sum_{s_2^{\,\prime}=-\infty}^{\infty}
\sum_{p=0}^l \sum_{q=-\infty}^{\infty} \sum_{s_1=-\infty}^{\infty}\sum_{s_2=-\infty}^{\infty}
C_{l0p^{\,\prime}q^{\,\prime}s_1^{\,\prime}s_2^{\,\prime}} \, C_{l0pqs_1s_2} \, \beta_{l0pqs_1s_2} \, k_l(\beta_{l0pqs_1s_2})  \\
\times \sin([2l-2p^{\,\prime} -2p] \omega + [2l-2p^{\,\prime}+q^{\,\prime} - s_1^{\,\prime} -2p+q - s_1] {\cal{M}}
-s_2^{\,\prime} (\chi t +\varphi) - s_2 (\chi t +\varphi)
- \epsilon_l(\beta_{l0pqs_1s_2}))  \\
 + \frac{1}{2} \sum_{p^{\,\prime}=0}^l \sum_{q^{\,\prime}=-\infty}^{\infty} \sum_{s_1^{\,\prime}=-\infty}^{\infty}\sum_{s_2^{\,\prime}=-\infty}^{\infty}
\sum_{m=0}^l \sum_{p=0}^l \sum_{q=-\infty}^{\infty} \sum_{s_1=-\infty}^{\infty}\sum_{s_2=-\infty}^{\infty}
C_{lmp^{\,\prime}q^{\,\prime}s_1^{\,\prime}s_2^{\,\prime}} \, C_{lmpqs_1s_2} \, \beta_{lmpqs_1s_2} \,k_l(\beta_{lmpqs_1s_2})  \\
\times \sin([-2p^{\,\prime} +2p] \omega + [-2p^{\,\prime}+q^{\,\prime} - s_1^{\,\prime} +2p-q + s_1] {\cal{M}}
-s_2^{\,\prime} (\chi t +\varphi) + s_2 (\chi t +\varphi)
+ \epsilon_l(\beta_{lmpqs_1s_2})) \\
+...
 \label{}
 \end{split}
 \end{equation}
 where the ellipsis stand for the terms that will vanish after averaging over the longitude $\lambda$; the quantities $\,C_{lmpqs_1s_2}\,$ are given by
 \ba
 C_{lmpqs_1s_2}\,=\,-\,\frac{GM^*}{a} \bigg( \frac{R}{a} \bigg)^l \frac{(l-m)!}{(l+m)!} (2-\delta_{0m}) P_{lm}(\sin \phi) F_{lmp} (i) G_{lpq}(e) J_{s_1} (m {\cal{A}}_1)
 J_{s_2} (m {\cal{A}})\,\;,\qquad
 \ea
 while $\,\beta{lmpqs_1s_2}\,$ is the six-index tidal Fourier mode given by the expression (\ref{six}).

 Averaging over the motion of the pericentre and over the mean motion, we observe that in the first sum only the terms with $\,p^{\,\prime} = l-p\,$ and $\,s_1^{\,\prime} = q^{\,\prime} + q - s_1\,$ survive, while in the second sum only the terms with $\,p^{\,\prime} = p\,$ and $\,s_1^{\,\prime} = q^{\,\prime}-q+s_1\,$ survive. Averaging independently over the period of the free libration, we see that in the first sum only the terms with $\,s_2^{\,\prime} =  - s_2\,$ stay, while in the second sum only the terms with $\,s_2^{\,\prime} = s_2\,$ are left.

 Continuing like in Appendix E.3, we shall arrive at a sum of terms containing all the products $\;J_{q^{\,\prime}-q+s_1}(m {\cal{A}}_1)\,J_{s_1}(m {\cal{A}}_1)\,J^2_{s_2}(m {\cal{A}})\;$. If however we agree to keep only the terms which are, at most, quadratic in the libration magnitudes, then the cross terms may be neglected and we shall be left only with the terms from $\;J_{q^{\,\prime}-q+s_1}(m {\cal{A}}_1)\,J_{s_1}(m {\cal{A}}_1)\;$ and those containing $\;J^2_{s_2}(m {\cal{A}})\;$. In other words, in the said approximation the total power will consist of two separate groups of terms: the terms due to the forced libration (with $\,s_2=0\,$ and $\,\beta_{lmpqs_10}\,$) and the terms due to the free libration (with $\,s_1=0\,$ and $\,\beta_{lmpq0s_2}\,$). In practical terms, this means that in the said approximation the total power will be simply a sum of the forced-libration power (\ref{terra}), with the five-indexed tidal mode $\,\beta_{lmpqs}\,$ given by expression (\ref{119}), and the free-libration power (\ref{dud}), with the five-indexed tidal mode $\,\beta_{lmpqs}\,$ given by expression (\ref{11119}).

 \end{document}